\pdfoutput=1
\documentclass[11pt,twoside,a4paper,cmspaper,final,collab]{cms-tdr}
\def\svnVersion{1c01e0b}\def\svnDate{2024/10/21}\def\cmsCernNoTag{CERN-EP-2024-067}\def\cmsCernDate{\today}\def\cmsMessage{Published in Physical Review D as \href{http://dx.doi.org/10.1103/PhysRevD.110.072012}{\doi{10.1103/PhysRevD.110.072012}.}}
\begin{document}\cmsNoteHeader{B2G-19-001}

\newlength\cmsFigWidth
\newlength\cmsTabSkip\setlength{\cmsTabSkip}{1ex}
\ifthenelse{\boolean{cms@external}}{\setlength\cmsFigWidth{0.49\textwidth}}{\setlength\cmsFigWidth{0.65\textwidth}}
\ifthenelse{\boolean{cms@external}}{\providecommand{\cmsTable}[1]{#1}}{\providecommand{\cmsTable}[1]{\resizebox{\textwidth}{!}{#1}}}
\DeclareRobustCommand{\tprime}{{\HepParticle{T}{}{}}\Xspace}
\DeclareRobustCommand{\bprime}{{\HepParticle{B}{}{}}\Xspace}
\newcommand{\tH}{\ensuremath{\PQt\PH}\xspace}
\newcommand{\tZ}{\ensuremath{\PQt\PZ}\xspace}
\newcommand{\Hbb}{\ensuremath{\PH\to\bbbar}\xspace}
\newcommand{\tHbq}{\ensuremath{\PQt\PH\PQb\PQq}\xspace}
\newcommand{\tZbq}{\ensuremath{\PQt\PZ\PQb\PQq}\xspace}
\newcommand{\tprimetobW}{\ensuremath{\tprime\to\PQb\PW}\xspace}
\newcommand{\ttjets}{\ensuremath{\ttbar}+jets\xspace}
\newcommand{\Tbq}{\ensuremath{\Pp\Pp\to\tprime\PQb\PQq}\xspace}

\cmsNoteHeader{B2G-19-001}

\title{Search for production of a single vector-like quark decaying to \texorpdfstring{\tH}{tH} or \texorpdfstring{\tZ}{tZ} in the all-hadronic final state in \texorpdfstring{$\Pp\Pp$}{pp} collisions at \texorpdfstring{$\sqrt{s} = 13\TeV$}{sqrt(s) = 13 TeV}}

\date{\today}

\abstract{
    A search for electroweak production of a single vector-like \tprime quark in association with a bottom (\PQb) quark in the all-hadronic decay channel is presented. This search uses proton-proton collision data at $\sqrt{s} = 13\TeV$ collected by the CMS experiment at the CERN LHC during 2016--2018, corresponding to an integrated luminosity of 138\fbinv. The \tprime quark is assumed to have charge $2/3$ and decay to a top (\PQt) quark and a Higgs (\PH) or \PZ boson. Hadronic decays of the \PQt quark and the \PH or \PZ boson are reconstructed from the kinematic properties of jets, including those containing \PQb hadrons. No deviation from the standard model prediction is observed in the reconstructed \tH and \tZ invariant mass distributions. The 95\% confidence level upper limits on the product of the production cross section and branching fraction of a \tprime quark produced in association with a \PQb quark and decaying via \tH or \tZ range from 1260 to 68\unit{fb} for \tprime quark masses of 600--1200\GeV.
}

\hypersetup{
    pdfauthor={CMS Collaboration},
    pdftitle={Search for production of single vector-like quarks decaying to tH or tZ in the all-hadronic final state in pp collisions at sqrt(s) = 13 TeV},
    pdfsubject={CMS},
    pdfkeywords={CMS, Tprime, vector-like quark}
}

\maketitle

\section{Introduction}
\label{sec:intro}

Vector-like quarks (VLQs) appear in many extensions of the standard model (SM) as they provide a solution to the hierarchy problem by protecting the Higgs (\PH) boson mass from large radiative corrections~\cite{Okada:2012gy,Aguilar-Saavedra:2013qpa, DeSimone:2012fs, Buchkremer:2013bha}.
They are assumed to mix with the third-generation SM quarks, and in contrast to SM chiral quarks, have both vector and axial vector-like couplings.
These nonchiral couplings allow VLQs to avoid the strong experimental constraints, which now exclude additional generations of chiral quarks~\cite{Djouadi:2012ae,Lenz:2013iha,Chatrchyan:2012fp}.

\begin{figure}[ht]
    \centering
    \includegraphics[width=0.49\textwidth,height=0.25\textheight]{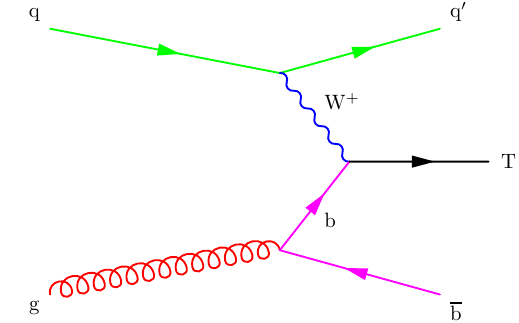}
    \caption{Example of Feynman diagram for electroweak production of a vector-like \tprime quark.}
    \label{fig:diagrams}
\end{figure}

The VLQs can either be produced in pairs via the strong interaction or singly via the electroweak interactions, in association with additional quarks.
Figure~\ref{fig:diagrams} shows a possible electroweak production diagram for a vector-like top quark partner \tprime of charge $2/3$.
While the pair production cross section depends only on the \tprime quark mass, the electroweak production of a single \tprime quark depends on the initial-state coupling.
These couplings are typically constrained to small values by precision measurements~\cite{Aguilar-Saavedra:2013qpa}, thus resulting in low production cross sections.
However, in some models, such as a composite Higgs scenario~\cite{Vignaroli:2012nf}, a cross section of up to two orders of magnitude larger is possible.

The \tprime quark may decay into a bottom (\PQb) quark and a \PW boson ($\PQb\PW$), a top (\PQt) quark and a \PZ boson (\tZ), or a top quark and a Higgs boson (\tH).
The branching fractions are model dependent.
For the singlet model presented in Ref.~\cite{AguilarSaavedra:2009es}, the branching fractions are 50\% for decays to $\PQb\PW$ and 25\% each for decays to \tZ or \tH.
In contrast, if the \tprime quark occurs in a doublet with a vector-like bottom quark partner \bprime of charge $-1/3$, its branching fractions is 50\% for decays to either \tZ or \tH.

We search for electroweak production of a \tprime quark in association with a \PQb quark, which requires a nonzero $\tprime\PW\PQb$ coupling for production.
We consider a \tprime quark with a narrow width ($\Gamma/M \lesssim 1\%$) in the singlet model with a 50\% branching fraction to $\PQb\PW$ and 25\% branching fractions each to \tZ and \tH.

The ATLAS and CMS Collaborations have reported many searches for pair production of narrow-width VLQs in various final states using proton-proton ($\Pp\Pp$) collisions at $\sqrt{s} = 13\TeV$~\cite{Sirunyan:2017pks,ATLAS:2018uky,ATLAS:2022tla,CMS:2022fck,ATLAS:2022hnn}.
Both Collaborations have set constraints on electroweak production of VLQs~\cite{ATLAS:2018dyh,ATLAS:2018cjd,ATLAS:2022ozf}.
The CMS Collaboration has searched for \tprimetobW~\cite{Sirunyan:2017tfc}, \tZ~\cite{Sirunyan:2017ynj,CMS:2022yxp}, and \tH~\cite{Khachatryan:2016vph,Sirunyan:2016ipo}, primarily focusing on signatures with a large Lorentz boost of the \tprime quark decay products.

In the fully hadronic final state, a signature of single \tprime quark production is a resonant peak in the reconstructed \PQt + \bbbar mass distribution, with the \bbbar pair originating from the decay of either a \PH or \PZ boson.
Events with five jets consistent with hadronic decays of a top quark and a \PH or \PZ boson are selected.
The reconstruction of the \PH or \PZ boson candidate is achieved using \PQb tagging techniques and dedicated kinematic selections.
This signature is particularly effective for \tprime masses when the final state jets can be resolved. For masses above 1000\GeV, some decay products may merge into a single jet, requiring targeted reconstruction algorithms not considered in this analysis.
The search presented in this paper uses $\Pp\Pp$ collision data collected at $\sqrt{s} = 13\TeV$ over the 2016, 2017, and 2018, and corresponds to an integrated luminosity of 138\fbinv.
The analysis is focused on \tprime quark masses between 600--1200\GeV and extends results reported by CMS in Ref.~\cite{Sirunyan:2019xeh} that are based on the 2016 data.

The paper is organized as follows.
In Section~\ref{sec:detector} the CMS detector and event reconstruction are described, modeling of signal and background processes is discussed in Section~\ref{sec:sbmodeling}.
Methods for reconstructing the \tprime quark mass and event selection are presented in Section~\ref{sec:reconstruction}, and background estimation method is described in Section~\ref{sec:fh-BkgEst}.
Section~\ref{sec:systematics} discusses the systematic uncertainties considered, results are shown in Section~\ref{sec:YearResults} and a summary is given in Section~\ref{sec:summary}.
\section{The CMS detector and event reconstruction}
\label{sec:detector}

The central feature of the CMS apparatus is a superconducting solenoid of 6\unit{m} internal diameter, providing a magnetic field of 3.8\unit{T}.
Within the solenoid volume are a silicon pixel and strip tracker, a lead tungstate crystal electromagnetic calorimeter (ECAL), and a brass and scintillator hadron calorimeter (HCAL), each composed of a barrel and two endcap sections.
Forward calorimeters extend the pseudorapidity ($\eta$) coverage provided by the barrel and endcap detectors.
Muons are detected in gas-ionization chambers embedded in the steel flux-return yoke outside the solenoid.

The silicon tracker used in 2016 measured charged particles within the range $\abs{\eta} < 2.5$.
For nonisolated particles of transverse momentum $1 < \pt < 10\GeV$ and $\abs{\eta} < 1.4$, the track resolutions were typically 1.5\% in \pt and 25--90 (45--150)\mum in the transverse (longitudinal) impact parameter~\cite{CMS:2014pgm}.
At the start of 2017, a new pixel detector was installed~\cite{Phase1Pixel}; the upgraded tracker measured particles up to $\abs{\eta} < 3.0$ with typical resolutions of 1.5\% in \pt and 20--75\mum in the transverse impact parameter~\cite{DP-2020-049} for nonisolated particles of $1 < \pt < 10\GeV$.

The electron momentum is estimated by combining the energy measurement in the ECAL with the momentum measurement in the tracker.
The momentum resolution for electrons with $\pt \approx 45\GeV$ from $\PZ \to \Pe \Pe$ decays ranges from 1.6 to 5\%.
It is generally better in the barrel region than in the endcaps, and also depends on the bremsstrahlung energy emitted by the electron as it traverses the material in front of the ECAL~\cite{CMS:2020uim,CMS-DP-2020-021}.
In the region $\abs{\eta} < 1.74$, the HCAL cells have widths of 0.087 in pseudorapidity and 0.087 in azimuth ($\phi$).
In the $\eta$-$\phi$ plane and for $\abs{\eta} < 1.48$, the HCAL cells map on to $5 \times 5$ arrays of ECAL crystals to form calorimeter towers projecting radially outwards from close to the nominal interaction point.
For $\abs{\eta} > 1.74$, the coverage of the towers increases progressively to a maximum of 0.174 in $\Delta \eta$ and $\Delta \phi$.
Within each tower, the energy deposits in ECAL and HCAL cells are summed to define the calorimeter tower energies, subsequently used to provide the energies and directions of hadronic jets.

A more detailed description of the CMS detector, together with a definition of the coordinate system used and the relevant kinematic variables, can be found in Ref.~\cite{Chatrchyan:2008zzk}.

Events of interest are selected using a two-tiered trigger system~\cite{Khachatryan:2016bia}.
The first level, composed of custom hardware processors, uses information from the calorimeters and muon detectors to select events at a rate of around 100\unit{kHz} within a fixed time interval of about 4\mus.
The second level, known as the high-level trigger, consists of a farm of processors running a version of the full event reconstruction software optimized for fast processing, and reduces the event rate to around 1\unit{kHz} before data storage.

The primary $\Pp\Pp$ interaction vertex is taken to be the vertex corresponding to the hardest scattering in the event, evaluated using tracking information alone, as described in Section 9.4.1 of Ref.~\cite{CMS-TDR-15-02}.

The particle-flow (PF) algorithm~\cite{Sirunyan:2017ulk} aims to reconstruct and identify each individual particle (PF candidate) in an event, with an optimized combination of information from the various elements of the CMS detector.
The energy of photons is obtained from the ECAL measurement.
The energy of electrons is determined from a combination of the electron momentum at the primary interaction vertex as determined by the tracker, the energy of the corresponding ECAL cluster, and the energy sum of all bremsstrahlung photons spatially compatible with originating from the electron track.
The energy of muons is obtained from the curvature of the corresponding track.
The energy of charged hadrons is determined from a combination of their momentum measured in the tracker and the matching ECAL and HCAL energy deposits, corrected for the response function of the calorimeters to hadronic showers.
Finally, the energy of neutral hadrons is obtained from the corresponding corrected ECAL and HCAL energies.

The current search targets hadronic decays of the vector-like \tprime quark that result in at least five jets in the final state. Jets are reconstructed offline from the energy deposits in the calorimeter towers, clustered using the anti-\kt algorithm~\cite{Cacciari:2008gp, Cacciari:2011ma} with a distance parameter of 0.4.
In this process, the contribution from each calorimeter tower is assigned a momentum, the absolute value and the direction of which are given by the energy measured in the tower, and the coordinates of the tower.
The raw jet energy is obtained from the sum of the tower energies, and the raw jet momentum by the vectorial sum of the tower momenta, which results in a nonzero jet mass.
The raw jet energies are then corrected to establish a uniform relative response of the calorimeter in $\eta$ and a calibrated absolute response in \pt.
Additional $\Pp\Pp$ interactions within the same or nearby bunch crossings (pileup) can contribute additional tracks and calorimetric energy depositions, increasing the apparent jet momentum.
To mitigate this effect, tracks identified to be originating from pileup vertices are discarded, and an offset correction is applied to correct for remaining contributions~\cite{CMS:2020ebo}.
Jet energy corrections are derived from simulation studies so that the average measured energy of jets becomes identical to that of particle-level jets.
In situ measurements of the momentum balance in dijet, $\text{photon} + \text{jet}$, $\PZ + \text{jet}$, and multijet events are used to determine any residual differences between the jet energy scale in data and in simulation, and appropriate corrections are made~\cite{CMS:2016lmd}.
Additional selection criteria are applied to each jet to remove jets potentially dominated by instrumental effects or reconstruction failures~\cite{CMS:2020ebo}.
The jet energy resolution amounts typically to 15--20\% at 30\GeV, 10\% at 100\GeV, and 5\% at 1\TeV~\cite{CMS:2016lmd}.

Jets originating from the hadronization of a bottom quark can be tagged as \PQb jets.
The \textsc{DeepCSV} discriminator is used~\cite{BTV-16-002}.
It relies on a deep neural network trained to classify the different jet flavors using information from the tracks and secondary vertices.
Three working points~\cite{BTV-16-002} are defined: loose, medium, and tight, which correspond to a misidentification rate for light quark or gluon jets of 10, 1, and 0.1\%, respectively.
The efficiency for these working points, as measured in simulated samples of top quark-antiquark (\ttbar) production, ranges from around 80\% for the loose working point to around 45\% for the tight one.
\section{Modeling of signals and backgrounds}
\label{sec:sbmodeling}

Simulated samples for the signal process, \Tbq with the \tprime quark decaying to a \PQt quark and either a Higgs or a \PZ boson, are generated at leading order (LO) using the Monte Carlo (MC) event generator \MGvATNLO~\cite{MG5_aMCNLO} 2.2.2 (for data recorded in 2016) and 2.4.2 (for data recorded in 2017--2018).
The \tprime quark is assumed to have a width of 10\GeV, which is small compared to the experimental resolution, and left-handed chirality.
The mass of the top quark is set to 172.5\GeV and the mass of the Higgs boson to 125\GeV.
Both particles decay inclusively, with $\mathcal{B}(\Hbb)$ set to 58\%~\cite{PDG2022}.

The \MGvATNLO generator is interfaced with \PYTHIA~8.212~\cite{Sjostrand:2014zea} for the description of parton fragmentation and hadronization, using CUETP8M1~\cite{Khachatryan:2015pea} (2016) and CP5~\cite{CMS-GEN-17-001} (2017 and 2018) underlying event tunes.
Parton distribution function (PDF) sets used are LO NNPDF~3.0~\cite{Ball:2014uwa} (2016) and next-to-next-to-LO NNPDF~3.1~\cite{Ball:2017nwa} (2017 and 2018).
Simulation of the CMS detector is based on \GEANTfour~\cite{Agostinelli:2002hh}.
Simulated minimum-bias events are mixed with the hard interactions in simulated events to reproduce the effect of pileup.
Finally, all simulated events are reweighted such that the distribution of the number of collisions per bunch crossing matches the one observed in data, with an average of approximately 23 (32)~\cite{CMS-LUM-17-003, CMS-PAS-LUM-17-004, CMS-PAS-LUM-18-002} in 2016 (2017--2018).

The \ttbar events that are simulated with {\POWHEG~2.0}~\cite{1126-6708-2004-11-040,1126-6708-2007-11-070,Alioli2010,1126-6708-2007-09-126}
are used to cross check details of the analysis. The same generator is used for
the single top quark production. Simulated events from other SM background processes are used for cross checking the analysis strategy.
The \MGvATNLO generator is used to simulate the SM Higgs boson production in association with top quarks (\tH, $\ttbar\PH$)~\cite{hartanto_higgs_2015}, or with a vector boson ($\PV\PH$)~\cite{luisoni_hwhz_2013}.
The background processes \PW/\PZ+jets are simulated with \MGvATNLO.
Diboson and other SM events composed uniquely of jets produced through the strong interaction, referred to as quantum chromodynamics (QCD) multijet events, are produced at the leading order with \PYTHIA8.205~\cite{Sjostrand:2014zea}.
The PDF sets are identical to the signal ones and both change with the year.

The analysis uses data directly to estimate the background.
\section{Event selection}
\label{sec:reconstruction}

The analysis targets the decay of a \tprime quark to a top quark and a Higgs or \PZ boson, where $\PQt\to\PQb\PW\to\PQb\cPq\cPq'$ and $\PH/\PZ\to\bbbar$.
The final state is composed of at least seven jets, where five of them arise from the \tprime quark decay, three of which are \PQb jets. As shown in Fig.~\ref{fig:diagrams}, when the \tprime quark is singly produced, two additional jets are present in the event.

Events are selected online using a combination (logical ``OR'') of hadronic trigger criteria.
These triggers require a scalar \pt sum of the jets (\HT) above a certain threshold: 400\GeV in 2016, and 300\GeV in 2017 and 2018.
For the 2016 data, this combination includes a trigger requiring at least six jets with $\pt > 30\GeV$, with at least two of them passing the online \PQb tagging criteria.
For the 2017 and 2018 data-taking years, two main trigger paths are used: one requiring at least four jets with three of them \PQb tagged, and a second requiring at least six jets with two of them \PQb tagged.
For these triggers, the jet \pt threshold was increased to 32\GeV.
The trigger efficiency with respect to the offline selection is measured in data and found to be about 97\% for 2016, 95\% for 2017, and 97\% for 2018 with a dedicated trigger path. The set of triggers records events with an invariant mass of around 300\GeV or above.

The invariant mass reconstructed from five jets is used as the main discriminating variable.
In the fully hadronic final state, the main background processes are multijet and $\ttbar$+jets events.
Signal events are expected to exhibit a peak in the five-jet invariant mass distribution at the resonance mass, while background events, in the absence of selection criteria, would have a smoothly falling shape.

Based on the online requirements, at least six jets with $\pt > 40\GeV$ and $\abs{\eta}<4.5$ are required offline.
As signal jets result from the decay of a high-mass resonance, the \pt thresholds are increased to 170, 130, and 80\GeV for the leading, second-leading, and third-leading jets, respectively.
In addition, at least three \PQb-tagged jets using the tight \textsc{DeepCSV} working point~\cite{BTV-16-002} are required among the jets with $\abs{\eta}<2.4$ (2.5) in 2016 (2017 and 2018) data.
The $\abs{\eta}$ range is extended in 2017 and 2018 data because of the increased coverage of the upgraded pixel detector~\cite{Phase1Pixel}.

\subsection{Identification of \texorpdfstring{\tprime}{T} quark candidates}
A multistep $\chi^2$ minimization algorithm is used to identify the best Higgs or \PZ boson and top quark candidates among all five-jet combinations as in Ref.~\cite{Sirunyan:2019xeh}.
Resonance mass peaks are expected for the \PH/\PZ/\PW boson and the top quark, which are examined to increase the efficiency of identifying the correct jet assignment.

The algorithm first minimizes $\chi^2_{\PH/\PZ}$, defined as
\begin{equation}
    \chi^{2}_{\PH/\PZ} = \left( \frac{m_{\PH/\PZ}^\text{meas} - \mu_{\PH/\PZ}^\mathrm{MC}}{\sigma_{\PH/\PZ}^\mathrm{MC}} \right)^2,
    \label{eq:fh-chi2H}
\end{equation}
where $m_{\PH/\PZ}^\text{meas}$ is the measured invariant mass of a given pair of \PQb-tagged jets and $\mu_{\PH/\PZ}^\mathrm{MC}$ and $\sigma_{\PH/\PZ}^\mathrm{MC}$ denote the expected mean and standard deviation mass values obtained from Gaussian fits to the distribution of the invariant mass computed from matched reconstructed jets in simulated signal samples.
The reconstructed jets are matched to the generated particles if the $\Delta R$ between them is less than 0.4.
The minimization is performed over all possible pairs of \PQb-tagged jets.
The chosen \PQb-tagged jets are removed from consideration for the subsequent steps.

The remaining jets are considered for forming a \PW boson candidate, consisting of two jets, and a top quark candidate, consisting of the \PW boson candidate and an additional \PQb-tagged jet.
For these candidates, $\chi^{2}_{\PW}$ and $\chi^{2}_{\PQt}$ are calculated as
\begin{equation}\begin{aligned}
        \chi^{2}_{\PW}  & = \left( \frac{m_{\PW}^\text{meas} - \mu_{\PW}^\mathrm{MC}}{\sigma_{\PW}^\mathrm{MC}} \right)^2,    \\
        \chi^{2}_{\PQt} & = \left( \frac{m_{\PQt}^\text{meas} - \mu_{\PQt}^\mathrm{MC}}{\sigma_{\PQt}^\mathrm{MC}} \right)^2,
    \end{aligned}\end{equation}
with the analogous definitions of $m^\text{meas}$, $\mu^\mathrm{MC}$, and $\sigma^\mathrm{MC}$ for each resonance.
Finally, the total $\chi^2$ function
\begin{equation}
    \chi^{2} = \chi^{2}_{\PH/\PZ} + \chi^{2}_{\PW} + \chi^{2}_{\PQt}
    \label{eq:fh-chi2Tot}
\end{equation}
is minimized using all possible jet combinations for the \PW boson and top quark candidates.
This multistep procedure improves the signal-to-background ratio for a signal with a \tprime mass of 700\GeV by 30\% compared to simply choosing the jet combination with the best total $\chi^2$.

The values used for the particle mass mean and standard deviation are given in Table~\ref{tab:ResolvedMCvalues} and vary slightly for each data-taking year.
Using masses obtained by fitting the signal MC samples, improves the signal efficiency by 1--3\% compared to using world-average values.

\begin{table}[h]
    \centering
    \topcaption{
        Mean and standard deviation values from a Gaussian fit of the \PH/\PZ/\PW boson and top quark mass distributions in the 700\GeV \tprime quark sample, requiring the jet kinematic criteria described above and matching to generated particles.
        All quantities are in units of \GeVns.
        The year-to-year variations are within the jet energy scale uncertainties.
        \label{tab:ResolvedMCvalues}
    }
    \begin{scotch}{lcccccc}
        & \multicolumn{2}{c}{2016} & \multicolumn{2}{c}{2017} & \multicolumn{2}{c}{2018}                                                                   \\
        Particle & $\mu^\mathrm{MC}$        & $\sigma^\mathrm{MC}$     & $\mu^\mathrm{MC}$        & $\sigma^\mathrm{MC}$ & $\mu^\mathrm{MC}$ & $\sigma^\mathrm{MC}$ \\ \hline
        \PH      & 121.9                    & 13.5                     & 118.9                    & 14.7                 & 120.2             & 14.3                 \\
        \PZ      & 90.9                     & 11.4                     & 89.2                     & 12.0                 & 90.9              & 11.3                 \\
        \PW      & 83.8                     & 10.9                     & 82.5                     & 12.6                 & 83.9              & 10.8                 \\
        \PQt    & 173.8                    & 16.0                     & 172.8                    & 18.9                 & 175.9             & 17.2                 \\
    \end{scotch}
\end{table}

\subsection{Mass categorization}
\label{sec:masscategorization}

We impose additional selection criteria to reduce the background contribution and ensure correct identification of the \PH/\PZ/\PW boson and top quark candidates.
In addition to the kinematic selection described above, the baseline selection includes the following criteria.
\begin{itemize}
    \item The total $\chi^2 < 15$.
    \item The reconstructed Higgs (\PZ) boson must have a mass larger (smaller) than 100\GeV.
          This ensures the two channels do not overlap.
    \item The ``second top quark mass'' is defined as the invariant mass of the \PH/\PZ candidate and the highest \pt jet that is not used to form the primary top quark candidate.
          The second top quark mass must be greater than 250\GeV.
\end{itemize}

The second top quark mass requirement removes a large fraction of \ttbar events while retaining nearly all signal events, improving the signal-to-\ttbar ratio by more than 10\%.
Although three \PQb-tagged jets are required in the signal region (SR) of this search, a charm (\PQc) quark from the \PW boson decay misidentified as a \PQb quark could form a \PH/\PZ boson candidate with the genuine \PQb quark from the other top quark decay.
The \PQc quark misidentification by the \PQb tagging algorithms, which is at the level of a few percent, introduces a large \ttbar background.
These events are more likely than signal events to contain a second top quark candidate with a mass near 172\GeV.

Further selection criteria, described below, are intended to improve the quality of the \PH/\PZ/\PW boson, and top quark candidates to reduce the background.
These criteria are chosen to be as model-independent as possible.

Different selection criteria are used for low- and high-mass \tprime quark candidates in the mass range 600--800 and above 800\GeV, respectively.
The variables used for both selections are the same and are identical to those used in an earlier version of the analysis~\cite{Sirunyan:2019xeh}.
The high-mass selection consists of simple fixed-threshold requirements on the variables, while the low-mass requirements are chosen to avoid distorting the five-jet invariant mass distribution and producing artificial peaks.

The SR with the highest signal-to-background ratio is the 3T one requiring three tight \PQb-tagged jets.
The 3M SR contains events with three medium b-tagged jets, but no events with three tight \PQb-tagged jets.
The 2M1L SR contains events with two medium and one loose \PQb-tagged jets, but no events with three medium \PQb-tagged jets.
These conditions, along with the 3T SR, create three mutually exclusive regions.
The 2M1L SR is enriched in background events, while the 3M region provides a transition region between background- and signal-enriched regions.
Both regions have kinematic distributions similar to the 3T region.

\subsubsection{High-mass \texorpdfstring{\tprime}{T} selection\label{sec:highmassselection}}
High efficiency for \tprime quark masses above 800\GeV is achieved through selection criteria that are independent of the reconstructed \tprime quark mass.

\begin{itemize}
    \item We require relative \HT, defined as the scalar \pt sum of the \PH/\PZ boson and \PQt quark candidates divided by the total \HT of the event (scalar \pt sum of all jets with $\pt > 30\GeV$ in the event) to be greater than 0.4.
          In single \tprime quark production most of the momentum is carried by the top quark and \PH/\PZ boson, therefore the relative \HT discriminates against \ttbar and QCD multijet events.

    \item We require $\max(\chi^{2}_{\PH/\PZ}, \chi^{2}_{\PW}, \chi^{2}_{\PQt}) < 3.0$.
          The individual $\chi^{2}$ values are required to have small values to decrease the rate of \PH/\PZ/\PW/\PQt misidentification.
          This criterion effectively imposes a mass window of no more than ${\pm}\sqrt{3}$ times the mass resolution for each candidate.

    \item $\Delta R < 1.1$ separation is required between $\PQb_{\PH/\PZ}$ and $\cPaqb_{\PH/\PZ}$ jets that form the \PH/\PZ candidate where $\Delta R = \sqrt{\smash[b]{(\Delta\eta)^2 + (\Delta\phi)^2}}$ and $\Delta \eta$ and $\Delta \phi$ are the separations in pseudorapidity and azimuthal angle, respectively. In signal events, the \PH/\PZ boson tends to have a moderate Lorentz boost ($\pt \lesssim 400\GeV$), such that the two \PQb-tagged jets are near each other without being completely merged.

    \item For the \tH (\tZ) channel, we require $\chi^{2}_{\PH/\PZ}<1.5$ (1.0) because almost all background events do not contain a genuine \PH/\PZ boson.
          This criterion is equivalent to a mass window of $\pm$17 (12)\GeV around the \PH (\PZ) boson mass.

    \item We require $\Delta R(j_{\PW}, {j'}_{\PW}) < 1.75$ between two jets $j_{\PW}, {j'}_{\PW}$ comprising the \PW boson candidate because most signal events contain \PW bosons with moderate Lorentz boost.

    \item We also require $\Delta R({\PQb}_{\PQt},\PW) < 1.2$ between the \PQb jet forming the top quark candidate and the \PW boson candidate because most signal events contain top quarks with moderate Lorentz boost.
\end{itemize}

The table~\ref{tab:fh-Regions} summarizes the selection criteria for the signal regions.
In the SR, called the 3T region, we additionally require three jets passing the tight \PQb tagging criterion of the \textsc{DeepCSV} algorithm, yielding 1256 events in data with 15.1 signal events expected for a \tprime quark mass of 900\GeV.
The estimated selection efficiencies for a 900\GeV \tprime quark signal and various background processes in the \tH channel and the 2016 data-taking year are presented in Table~\ref{tab:fh-effcutflow}.
The efficiencies vary slightly in the 2017 and 2018 data-taking years.

Such a selection has the drawback of sculpting the five-jet invariant mass distribution as shown in Fig.~\ref{fig:CutVsQuantile}.

\begin{table*}[htb]
    \centering
    \topcaption{
        Definitions of the signal and control regions for the high-mass selection.
        If the same selection is applied in all SRs and CRs, this is indicated by the $\div$ symbol in the latter.
        If no selection is applied, this is indicated by the \NA~symbol.
        The ``3T'', ``3M'', and ``2T1L/2M1L'' represent region with three tight, three medium, and two tight/medium plus one loose \PQb tagging requirements on the jets, respectively.
        \label{tab:fh-Regions}
    }\cmsTable{
        \begin{scotch}{lccc}
            Requirement                                            & SR 3T/3M/2M1L & CR Multijet 3T/3M/2M1L & CR \ttbar 2T1L/2M1L \\
            \hline
            \PQt quark candidate \PQb-tagged jet                 & \NA           & \NA                    & {T}/{M}             \\
            $\chi^2 $                                              & $<$15         & {$<$50}                & {$<$50}             \\
            Relative \HT                                           & $>$0.4        & $\div$                 & $\div$              \\
            $\max(\chi^2_{\PH/\PZ}, \chi^2_{\PW}, \chi^2_{\PQt})$ & $<$3          & {5--20}                & {3--5}              \\
            $\chi^{2} _{\PQt}$                                    & $<$3          & {$>$1}                 & {$<$1.5}            \\
            $\Delta R(\PQb_{\PH/\PZ}, \PAQb_{\PH/\PZ})$            & $<$1.1        & $\div$                 & {$<$1.5}            \\
            $\chi^{2} _{\PH/\PZ}$ (\tH channel)                    & $<$1.5        & $\div$                 & {$<$3.0}            \\
            $\chi^{2} _{\PH/\PZ}$ (\tZ channel)                    & $<$1.0        & $\div$                 & {$<$3.0}            \\
            $\Delta R(j_{\PW}, {j'}_{\PW})$                        & $<$1.75       & $\div$                 & $\div$              \\
            $\Delta R(b_{\PQt},\PW)$                              & $<$1.2        & $\div$                 & $\div$              \\
        \end{scotch}
    }
\end{table*}

\begin{table*}[htb]
    \centering
    \topcaption{
        Cumulative efficiencies of the high-mass selection criteria for signal and various simulated backgrounds in 2016 data.
        The first and last lines indicate the expected number of events normalized to an integrated luminosity of 35.9\fbinv.
        Only statistical uncertainties are reported.
        The ``Other backgrounds'' column includes {\PW/\PZ}+jets, single \PQt, $\ttbar\PH$, and $\ttbar\PZ$ background processes.
        The $\ttbar\PH$ and $\ttbar\PZ$ processes do not form a resonance and have production rates roughly the same as the signal rate.
        \label{tab:fh-effcutflow}
    }
    \begin{scotch}{lcccc}
        Cuts                                                   & \tprime (900\GeV) & Multijet       & \ttbar        & Other backgrounds \\
        \hline
        Baseline selection (35.9\fbinv)                    & $9.5 \pm 0.3$     & $9360 \pm 810$ & $2610 \pm 30$ & $353 \pm 23$      \\[\cmsTabSkip]
        Relative \HT                                           & 89.5\%            & 42.8\%         & 51.9\%        & 52.9\%            \\
        $\max(\chi^2_{\PH/\PZ}, \chi^2_{\PW}, \chi^2_{\PQt})$ & 55.9\%            & 14.1\%         & 25.1\%        & 21.8\%            \\
        $\Delta R(\PQb_{\PH/\PZ}, \cPaqb_{\PH/\PZ})$          & 51.1\%            & 7.5\%          & 11.9\%        & 8.9\%             \\
        $\chi^{2} _{\PH/\PZ}$                                  & 45.6\%            & 4.9\%          & 9.3\%         & 7.1\%             \\
        $\Delta R(j_{\PW}, {j'}_{\PW})$                        & 39.2\%            & 3.2\%          & 7.2\%         & 5.6\%             \\
        $\Delta R({\PQb}_{\PQt},\PW)$                        & 34.1\%            & 1.9\%          & 4.5\%         & 2.5\%             \\[\cmsTabSkip]
        High-mass selection (35.9\fbinv)                   & $3.2 \pm 0.2$     & $181 \pm 52$   & $117 \pm 6$   & $9.3 \pm 0.6$     \\
    \end{scotch}
\end{table*}

\subsubsection{Low-mass \texorpdfstring{\tprime}{T} selection\label{sec:lowmassselection}}
For the low-mass \tprime selection, we first identify the criteria listed above that modify the five-jet invariant mass distribution.
The two $\chi^2$ criteria do not affect the five-jet invariant mass distribution, therefore the corresponding selections are unchanged.

For the other selection criteria, the threshold values are modified depending on the five-jet invariant mass in order to preserve the shape of the five-jet invariant mass falling spectrum.
We consider the distributions of the variables $\Delta R(\PQb_{\PH/\PZ}, \cPaqb_{\PH/\PZ})$, $\Delta R(j_{\PW}, {j'}_{\PW})$, relative $\HT$, and $\Delta R({\PQb}_{\PQt},\PW)$ in bins of the five-jet invariant mass.
For each mass bin, a threshold is defined in order to keep a given fraction of events.
The resulting threshold values for a given variable are fit as a function of the five-jet invariant mass.
These fitted functions represent the mass-dependent criteria for each variable.
The quantile is chosen to provide the same signal efficiency as the high-mass selection for a \tprime quark mass of 700\GeV.

\begin{figure}[!hbtp]
    \centering
    \includegraphics[width=0.50\textwidth]{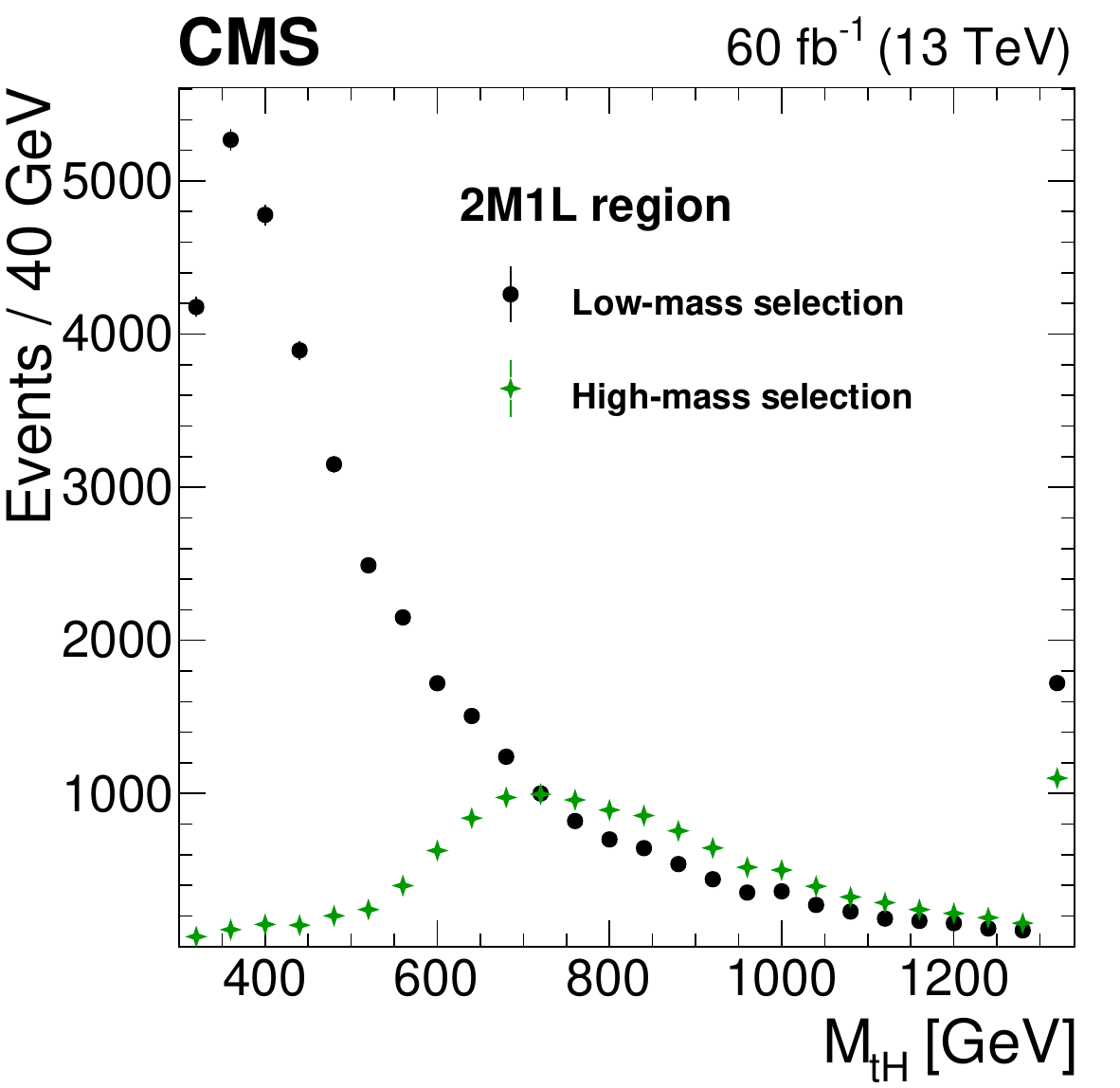}
    \caption{
        The five-jet invariant mass distribution in the 2M1L region after the high-mass (green crosses) and low-mass (black circles) selections in 2018 data.
        The rightmost bin of the distribution is an overflow bin.
        The low-mass selection results in a mass distribution that is smoothly falling, unlike the high-mass selection.
        The high-mass selection is more efficient for \tprime quark masses above 700\GeV by up to 25\% while maintaining a similar background level, as detailed in Section~\ref{sec:highmassselection}.
    }
    \label{fig:CutVsQuantile}
\end{figure}

The functions obtained from this procedure are polynomials of degree 5--7.
To mitigate the potentially large systematic uncertainty from the limited size of the MC sample of events with at least 3 tight \PQb-tagged jets, data are used directly.
The functions are derived using events from 2018 data in the two medium and one loose \PQb-tagged jet (2M1L) region, which excludes events with three medium \PQb-tagged jets.
To test for potential bias from the particular choice of function, we perform studies using alternative functional forms.
The presence of signal contamination at the signal strengths considered in this analysis would also have a negligible effect on the background model.
We find that neither small signal contamination nor the particular choice of function biases the selection.
The same functions are applied to the \tH and \tZ channels and to the other data-taking years.
As the low-mass selection always preserves the same fraction of events regardless of the five-jet invariant mass, for larger masses, the number of events is reduced compared to the high-mass selection corresponding to a lower signal efficiency.
Figure~\ref{fig:CutVsQuantile} compares the five-jet invariant mass distributions ($M_{\tH}$) after the low- and high-mass selections in the 2M1L region of the \tH channel in 2018 data.
With the low-mass selection, the five-jet invariant mass distribution is a smoothly falling spectrum above 400\GeV.

\section{Background estimation and validation}
\label{sec:fh-BkgEst}

While signal events follow a resonance in the five-jet invariant mass, SM background events, without any selection, form a smoothly falling distribution. The background distribution is inferred from regions with relaxed \PQb tagging requirements.
As described in Section~\ref{sec:masscategorization}, the high-mass selection sculpts the mass distribution.
The validation of the background estimation method is mainly conducted in this region as it has a more complex shape.

As the selection variables are not correlated with the \PQb tagging criteria, the shape of the five-jet invariant mass distribution for the SM backgrounds can be modeled from data in regions with loosely \PQb-tagged jets without introducing a significant bias.

The background-dominated regions are defined by relaxing the \PQb tagging requirements on three jets used to form the \tprime quark candidate.
With the large sample sizes in the 2M1L SR, the background distributions are determined with high statistical precision.

The background prediction is constructed bin by bin using a simultaneous binned maximum likelihood fit.
For each bin of the five-jet invariant mass, the total expected Poisson mean yield under the signal-plus-background hypothesis is composed of a background component and a signal component scaled by the signal strength $\mu$,
\begin{equation}\begin{aligned}
        \lambda_\mathrm{3T}   & = \lambda^{B}_\mathrm{3T}  + \mu \lambda^{S}_\mathrm{3T}      \\
        \lambda_\mathrm{3M}   & = \lambda^{B}_\mathrm{3M}  + \mu \lambda^{S}_\mathrm{3M}      \\
        \lambda_\mathrm{2M1L} & = \lambda^{B}_\mathrm{2M1L}  + \mu \lambda^{S}_\mathrm{2M1L},
        \label{eq:RooParametricHist}
    \end{aligned}\end{equation}
where $\lambda_\mathrm{X}$, $\lambda^{B}_\mathrm{X}$, and $\lambda^{S}_\mathrm{X}$ represent the expected total, background, and signal yields in a given bin and region X, respectively.
Each $\lambda^{S}_\mathrm{X}$ is determined from simulation.
The transfer functions connect the predictions in different regions,
\begin{equation}\begin{aligned}
        \lambda^{B}_\mathrm{3T} & = N_\mathrm{3M \to 3T} \,f_\mathrm{3M \to 3T} \, \lambda^{B}_\mathrm{3M}       \\
        \lambda^{B}_\mathrm{3M} & = N_\mathrm{2M1L \to 3M} \,f_\mathrm{2M1L \to 3M}\, \lambda^{B}_\mathrm{2M1L},
        \label{eq:Bkg}
    \end{aligned}\end{equation}
where $N_\mathrm{X \to Y}$ is an overall normalization factor, which corresponds to the ratio of the overall selection efficiency for region Y to that for region X and $f_\mathrm{X \to Y}$ is the value of the transfer function from region X to Y for a given bin correcting for small efficiency differences between the regions as described below.
The two normalization factors, signal strength, and $\lambda^{B}_\mathrm{2M1L}$ are simultaneously determined during the fit to data.
The fit procedure incorporates the systematic uncertainties as nuisance parameters described in the next section.

The transfer function accounts for \PQb tagging efficiencies for a given \PQb tagging working point to vary slightly depending on the \PQb-tagged jet kinematics.
To correctly model the background shape in a region with different \PQb tagging criteria, we reweight events based on the ratio of jets between loose and medium \PQb-tagged jets, and between medium and tight \PQb-tagged jets obtained in the multijet validation sample described below after requiring the baseline criteria. Since \pt and $\eta$ are highly correlated, we parametrize the ratios as functions of the total jet momentum and $\eta$.

The weight applied to each jet is calculated according to its kinematics, and the total event weight is the product of all the jet weights.
This event weight corrects for the difference in kinematics in \PQb tagging efficiency between the 3M and 3T (or 2T1L, corresponding to two tight and one loose \PQb-tagged jets) regions, and between the 2M1L and 3M regions.

\begin{figure*}[!hbtp]
    \centering
    \includegraphics[width=0.48\textwidth]{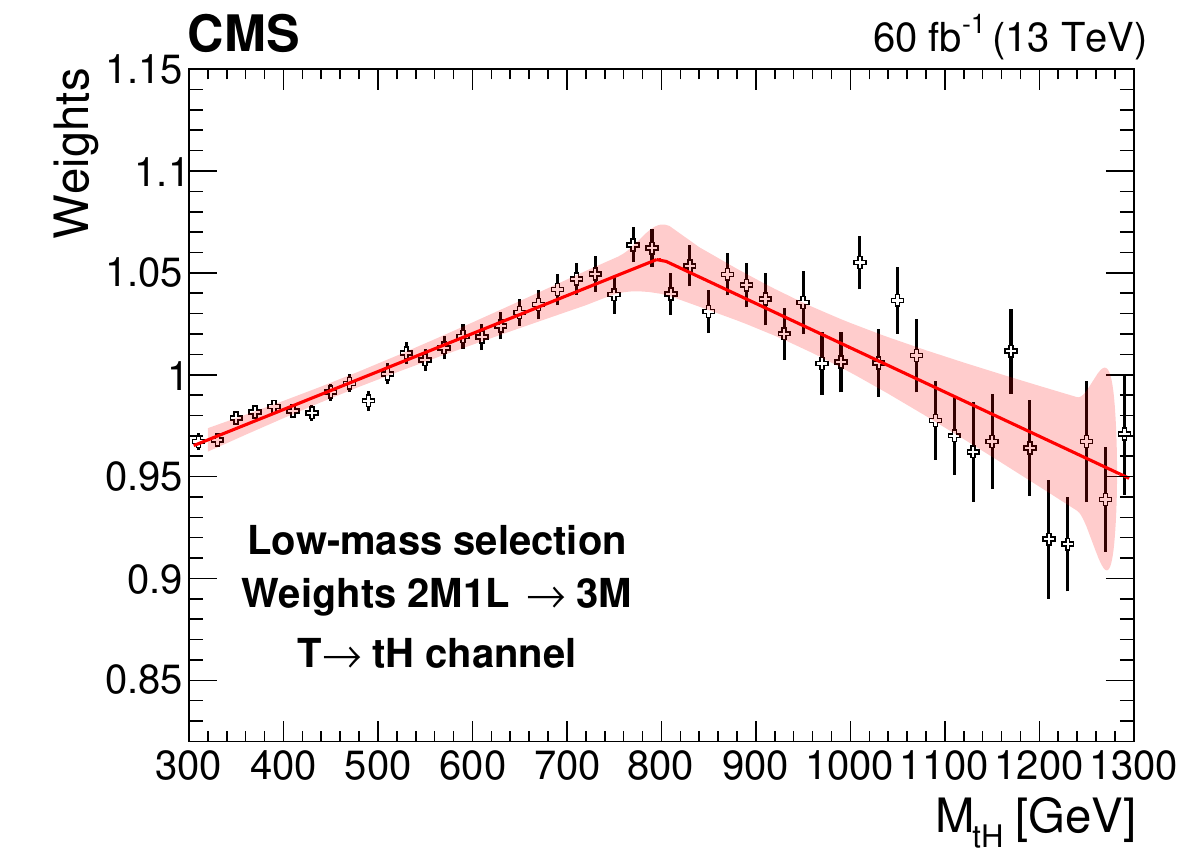}
    \includegraphics[width=0.48\textwidth]{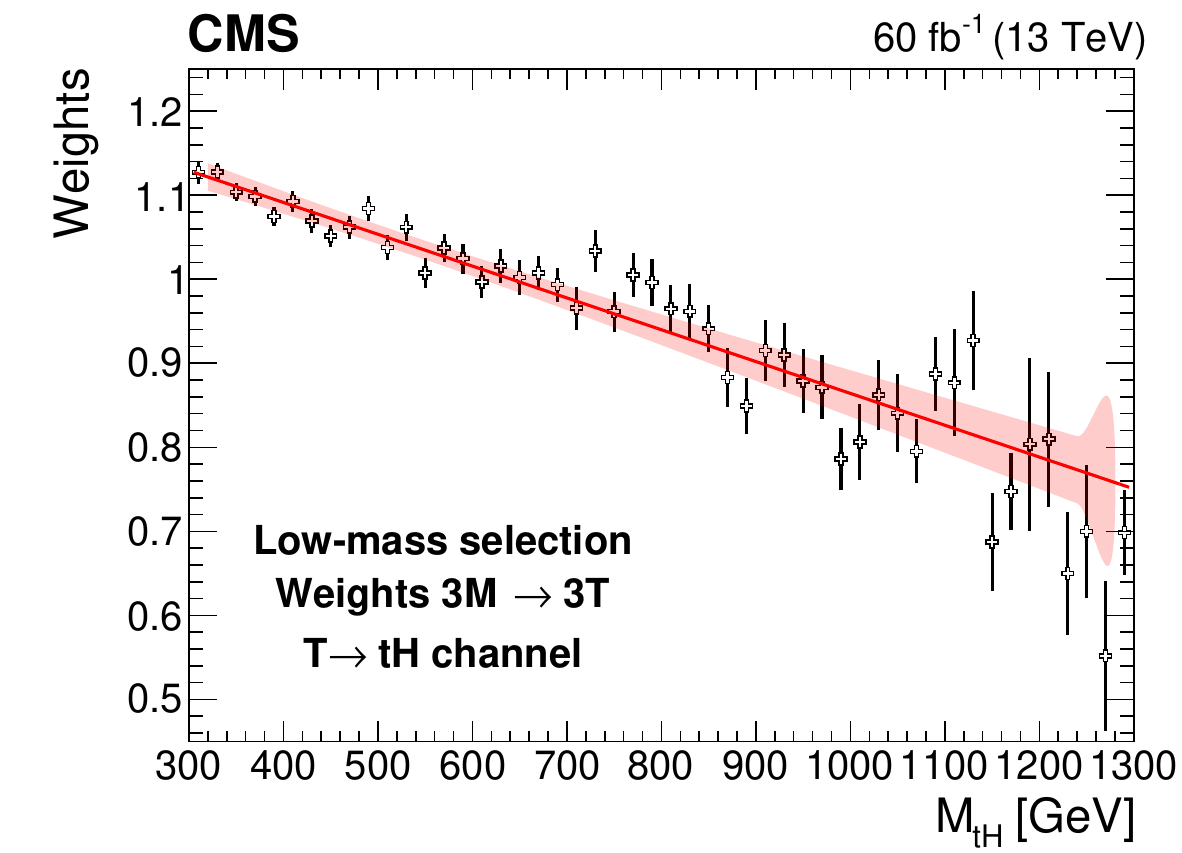}\\
    \includegraphics[width=0.48\textwidth]{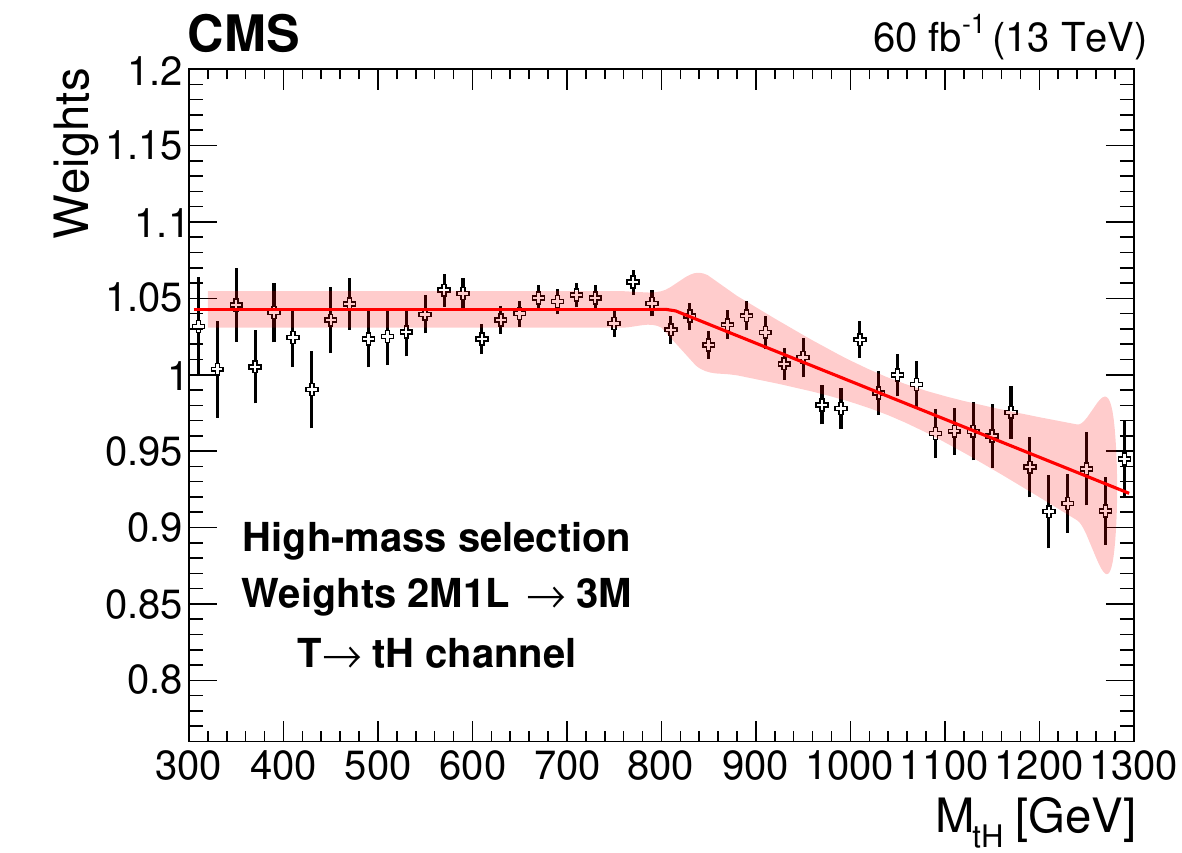}
    \includegraphics[width=0.48\textwidth]{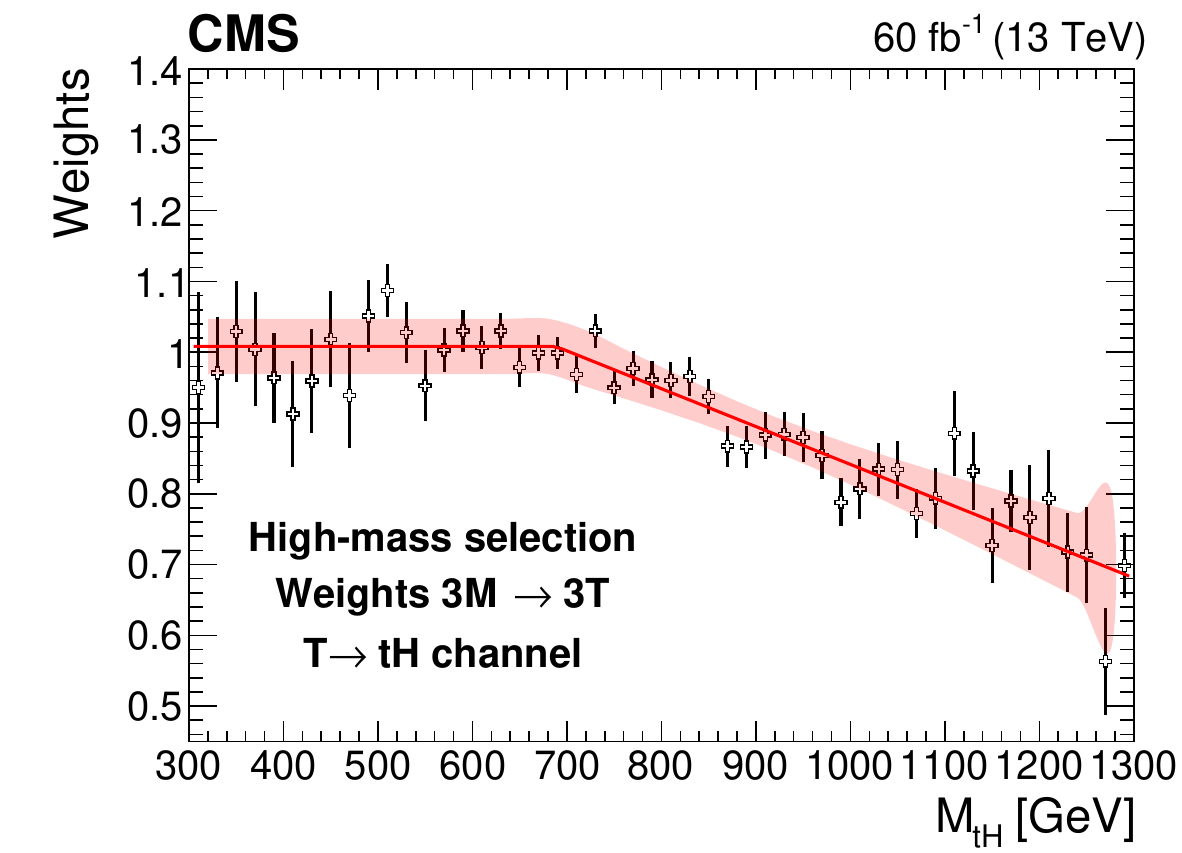}\\
    \caption{
        Weights from \PQb tagging ratios (open markers) as functions of the five-jet invariant mass in 2018 data for the low-mass (upper) and high-mass (lower) selections.
        The left graphs show weights connecting the 2M1L and 3M regions, and the right graphs show weights connecting the 3M and 3T regions.
        The red line corresponds to the central value of the transfer function and the shaded area represents the 95\% confidence level uncertainty band.
        For the low-mass (high-mass) analysis only signals with mass below (above) 800\GeV are tested, so primarily the lower (upper) part of the distribution contributes to the final result.
    }
    \label{fig:TF2018}
\end{figure*}

Transfer functions are derived by fitting a combination of two linear functions continuous at their connection to the weight distributions as functions of the five-jet invariant mass.
One transfer function connects the 3M and 3T regions, and a second transfer function connects the 2M1L and 3M regions.
Figure~\ref{fig:TF2018} shows both transfer functions for the low- and high-mass selections in the \tH channel for the 2018 data. The difference between the functions for the two selections is largely driven by the difference in the five-jet invariant mass shapes.
Similar functions are observed for the \tZ channel and the other data-taking years.
The range of the corrections is up to 5\% for the transfer function from the 2M1L to the 3M SRs and up to 30\% for the high-mass transfer function from 3M SR to the 3T SR.

The background shape estimation method is validated in control regions (CRs) enriched in multijet or \ttbar events where similar subsamples with different \PQb tagging criteria are defined.
Both samples include events with total $\chi^2 < 50$.
The multijet sample requires the maximum individual particle
$\chi^2$ to be in the range 5--20, while the \ttbar sample requires it to be in the range 3--5.
Both validation samples are mutually exclusive from the SRs where the maximum individual particle $\chi^2$ is less than 3.
In addition, the QCD multijet sample requires $\chi^2_{\PQt} > 1$ to reduce the number of \ttbar events, while the \ttbar sample requires $\chi^2_{\PQt} < 1.5$ along with an inverted $\chi^{2}_{\PH/\PZ}>1.5$ (1.0) requirement for the \tH (\tZ) channel.
The QCD multijet sample consists of approximately 80\% QCD multijet events and 20\% \ttbar events while the \ttbar sample contains approximately 70\% \ttbar events and 30\% QCD multijet events.
Both samples are further subdivided by the \PQb tag content using the same criteria as the background-dominated SRs described previously.

The \ttbar sample is split into two mutually exclusive 2M1L and 2T1L regions. In the 2T1L (2M1L) \ttbar CR, one of the tight (medium) \PQb-tagged jets must form part of the \PQt quark candidate.
Table~\ref{tab:fh-Regions} summarizes the definition of the signal region and control regions for the high-mass selection, highlighting the differences required to separate the samples.
In addition to these two CRs for the \tH and \tZ channels, more CRs are created by changing the $\mu_{\PQt}^\mathrm{MC}$ input to the $\chi^2$ variable: instead of setting it to the top quark mass, the value is fixed to either 140 or 250\GeV.
Regions analogous to those in Table~\ref{tab:fh-Regions} are then defined leading to five additional CRs.
The CRs with $\mu_{\PQt}^\mathrm{MC} = 250\GeV$ contain fewer events than the others.
The fraction of expected signal events is $\approx$3\% in the QCD multijet CR and $\approx$1\% in the \ttbar CR.

\begin{figure*}[!hbtp]
    \centering
    \includegraphics[width=0.32\textwidth]{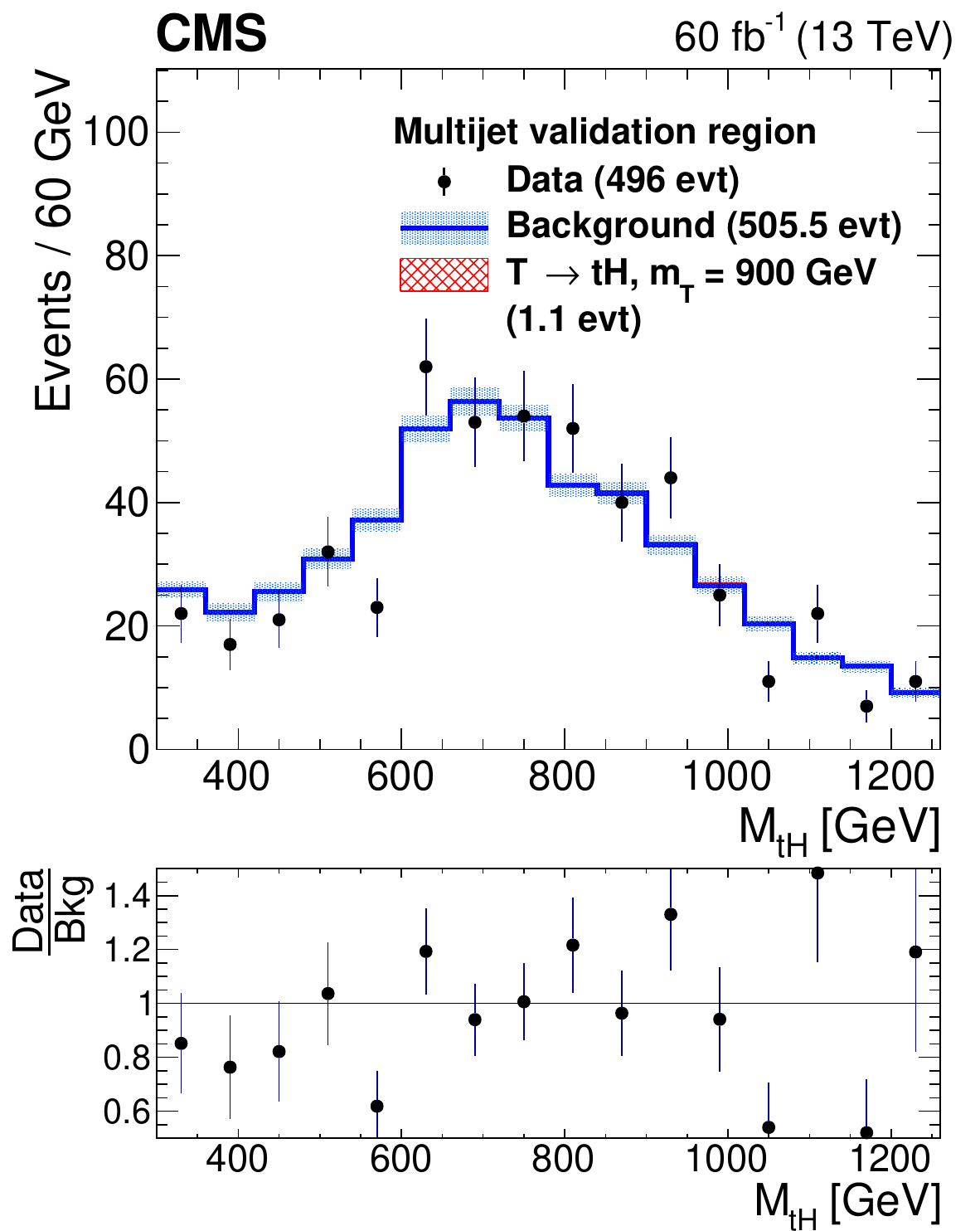}
    \includegraphics[width=0.32\textwidth]{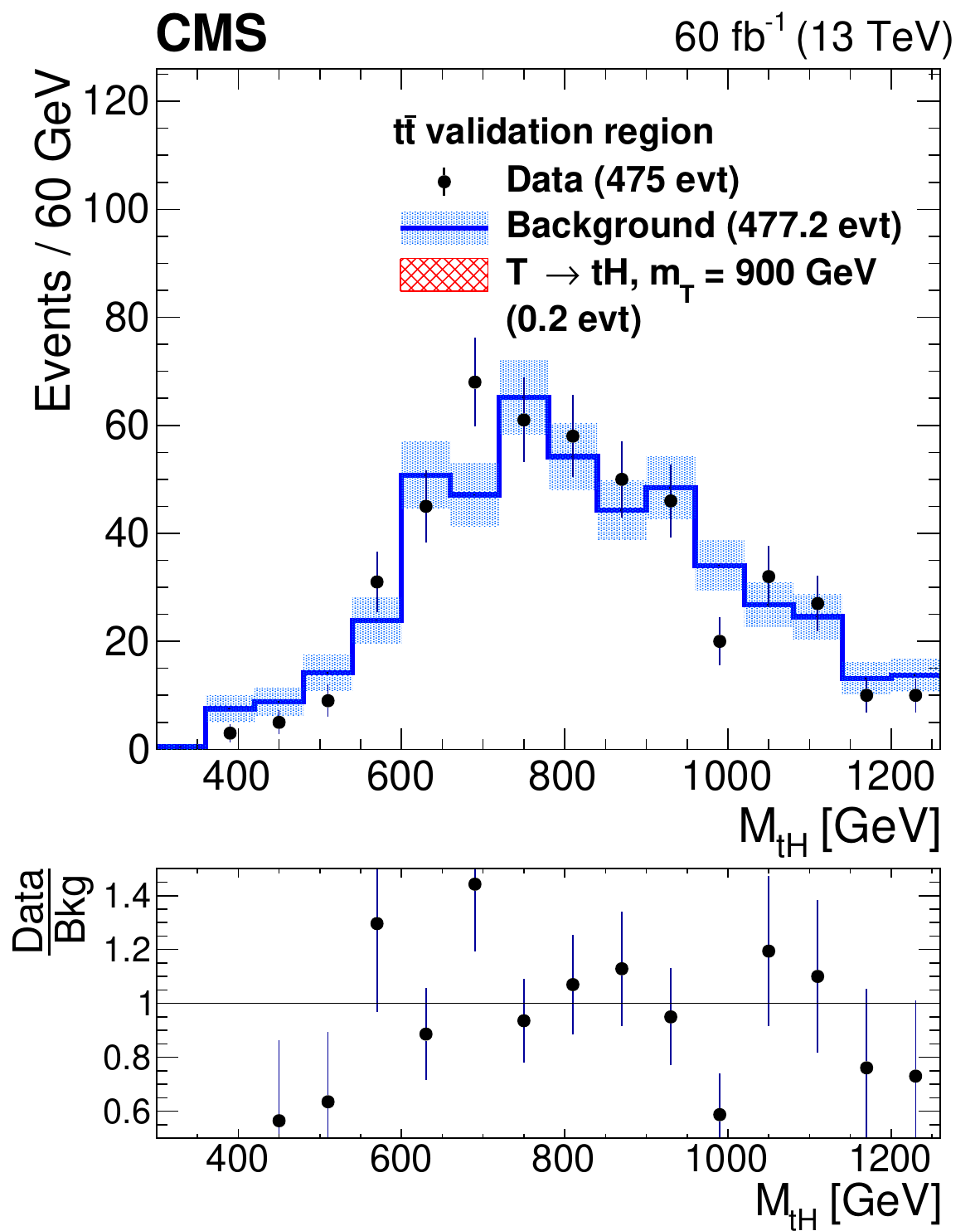}
    \includegraphics[width=0.32\textwidth]{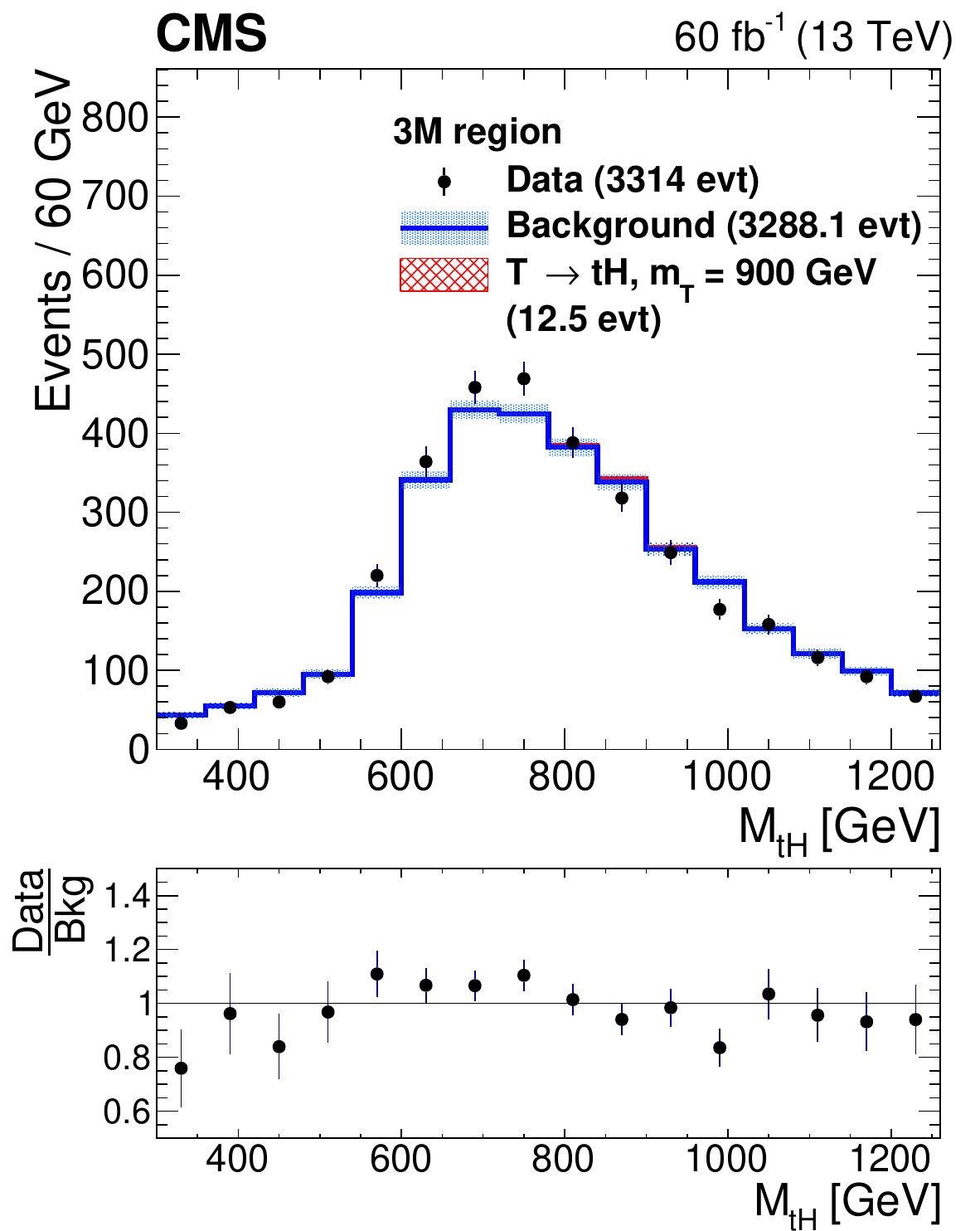}
    \caption{
        The five-jet invariant mass distribution in the \tH channel (black markers) after the high-mass selection in the QCD multijet 3T control region (left), the \ttbar 2T1L control region (middle), and the 3M signal region (right) for the 2018 data.
        The histograms are the corresponding reweighted 2M1L distributions.
        The background distribution is normalized to the number of entries in the data.
        The shaded area corresponds to the statistical uncertainties in the 2M1L control regions.
        A potential 900\GeV \tprime signal (red cross-hatched histogram) is added to the background histogram demonstrating a negligible contribution.
        Similar results are observed in the \tZ channel, and for the other years, but with slightly larger statistical uncertainties.
    }
    \label{fig:fh-FinalCR}
\end{figure*}

Validation of the background model is performed in the QCD multijet and \ttbar CRs. Events in the 2M1L CRs are reweighted to model the five-jet invariant mass distributions in the QCD multijet 3T or 3M CRs or \ttbar 2T1L CRs.
Kolmogorov-Smirnov tests~\cite{kolmogorov,smirnov1948,James:1019859} are used to compare the reweighted distributions with the data in the tighter regions, and they are found to be consistent with each other.
Figure~\ref{fig:fh-FinalCR} shows the five-jet invariant mass distributions in the CRs (QCD multijet 3T and \ttbar 2T1L) and the 3M SR in 2018 data, with the prediction from the corresponding reweighted 2M1L distributions.
A potential signal with a mass of 900\GeV is overlaid.
The shapes are similar in 2016 and 2017 data, and in the \tZ channel.
In total, 45 CRs have been examined.

\section{Systematic uncertainties}
\label{sec:systematics}

The systematic uncertainties in the analysis can be divided into two categories: those that affect the overall yields of the signal and background processes and those that affect the shape of the invariant mass distributions.
A 3 (5)\% uncertainty, uncorrelated between years, is assigned to the trigger efficiency for the 2016 (2017 and 2018) data.
The jet energy scale and resolution uncertainties affect the overall normalization and shapes of the signal mass distributions, leading to changes of a couple of percent in the mean and sigma of the Gaussian.
Correlations among the years are taken into account.
The $\PQb$ tagging efficiency scale factor uncertainties for jets are measured in multijet and \ttjets samples, separately for $\PQb$ jets and light-quark and gluon jets.
These uncertainties affect only the total event yields and are uncorrelated over the years.

The PDF uncertainties are evaluated using the {PDF4LHC} procedure~\cite{Butterworth:2015oua} and the NNPDF3.0 or NNPDF3.1 PDF sets, and are found to change the overall event yields by 0.5--1\%.
The uncertainty in the measurement of the integrated luminosity amounts to 2.5\%~\cite{CMS-LUM-17-003} for 2016, 2.3\%~\cite{CMS-PAS-LUM-17-004} for 2017, and 2.5\%~\cite{CMS-PAS-LUM-18-002} for 2018 and affects the normalization of all simulated processes.
The uncertainty in the mismodeling of the pileup is evaluated based on a 4.6\% variation of the $\Pp\Pp$ total inelastic cross section~\cite{ppInelastic2018} and affects the normalization of the simulated events.
This systematic uncertainty is correlated between the years.

The systematic uncertainties in the background estimation are assessed by multiplying the statistical uncertainties in the transfer function slope parameters by 4 for the transfer function from the 2M1L to the 3M SR and by 3 for the transfer function from the 3M to the 3T SR.
These factors cover the differences observed when the weights are computed in a \ttbar sample, enriched in heavy-flavor jets, compared to the multijet sample, populated mainly by light-flavor jets, as well as other uncertainties in the $\PQb$ tagging efficiency.
These uncertainties are uncorrelated across data-taking years because they are measured in statistically independent data sets.
\section{Results}
\label{sec:YearResults}

The shape of the signal is parametrized from simulation as a Gaussian distribution for each generated value of \tprime quark mass and separately for the low- and high-mass selections.
The background shape is determined from the 2M1L and 3M SRs using the corresponding transfer functions, while accounting for the potential presence of signal events.

Binned maximum likelihood fits are performed in the five-jet invariant mass distribution in each region including contributions from background and signal, scaled by the signal strength.
Systematic uncertainties, described in Section~\ref{sec:systematics}, are incorporated as nuisance parameters.
The fit is performed separately in the \tH and \tZ channels.
Up to 20\% reconstructed \tH events are misidentified as \tZ candidates. For such events, the shape of the signal remains Gaussian, except the five-jet invariant mass distribution is shifted to lower values compared to the \tH channel.
Combination of these events with the \tH is denoted as ``all-\tH'' channel.

\begin{figure*}[!hbtp]
    \centering
    \includegraphics[width=0.49\textwidth]{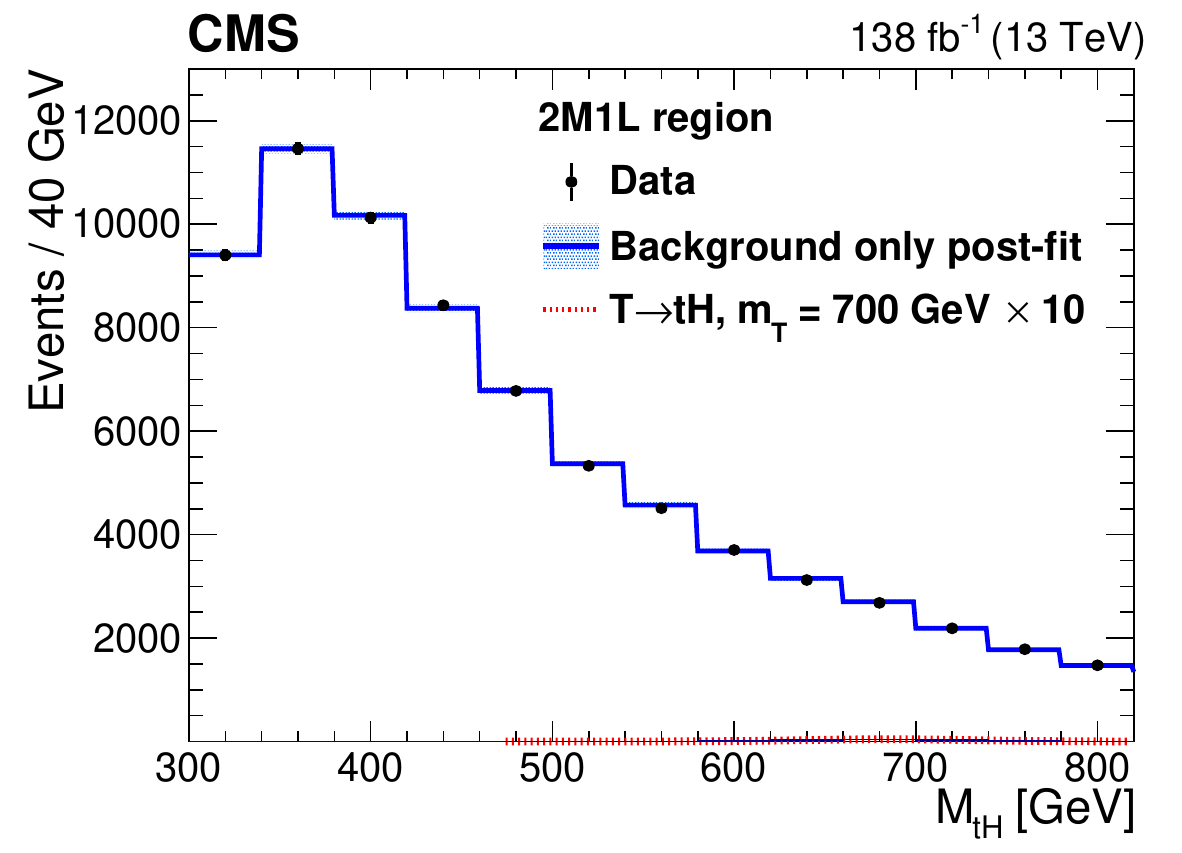}
    \includegraphics[width=0.49\textwidth]{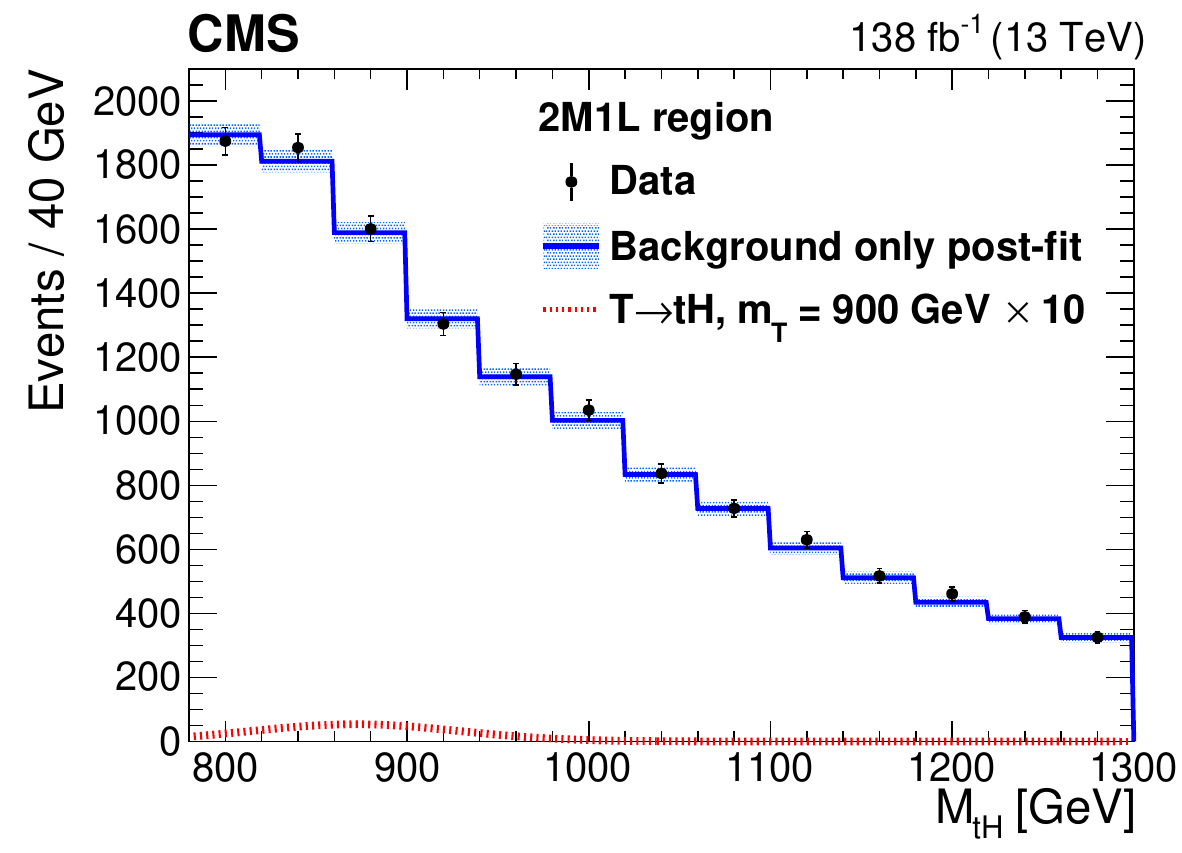}
    \includegraphics[width=0.49\textwidth]{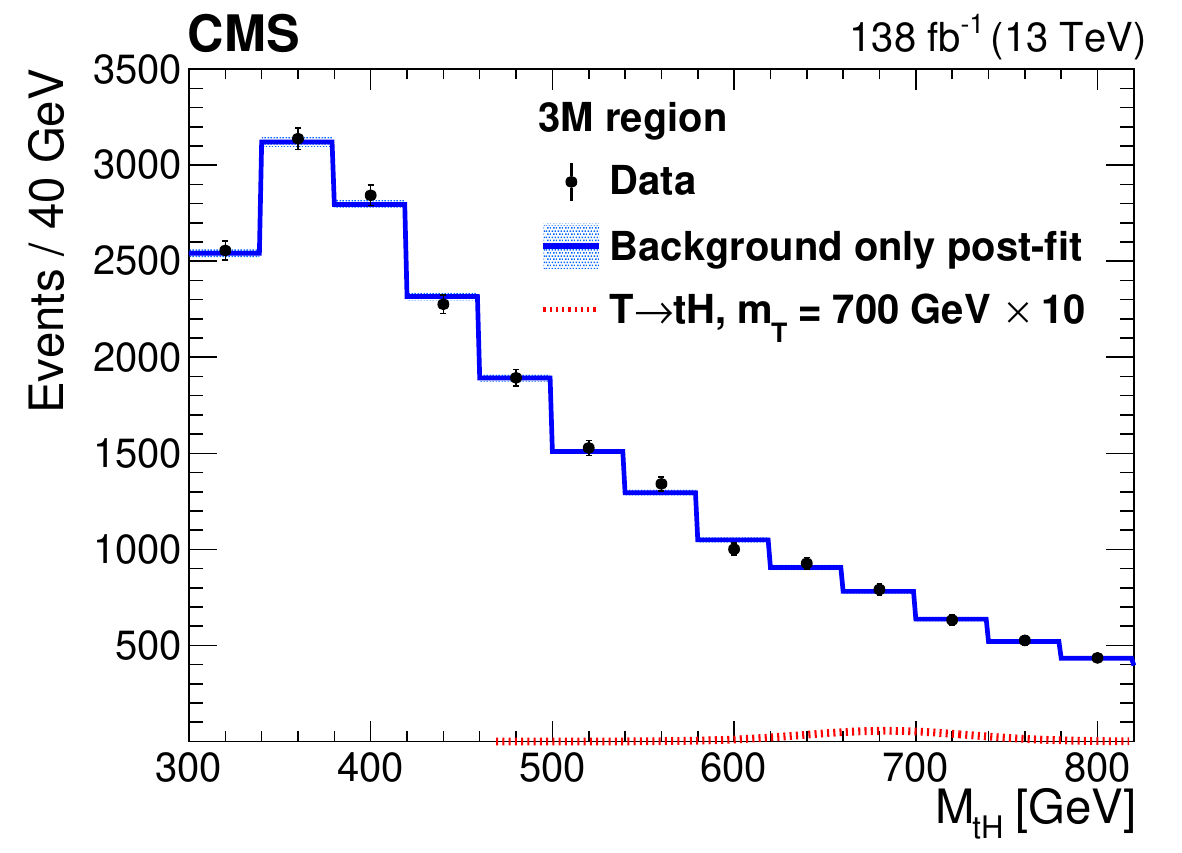}
    \includegraphics[width=0.49\textwidth]{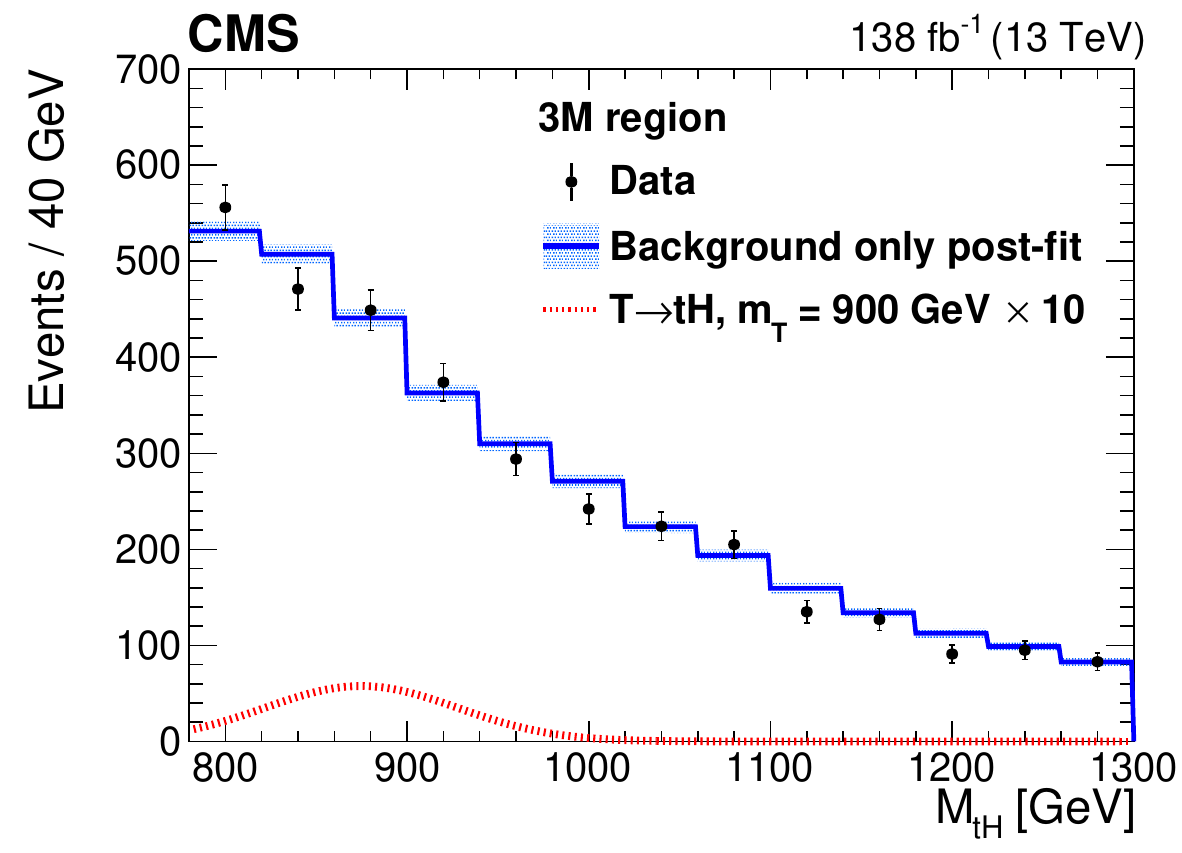}
    \includegraphics[width=0.49\textwidth]{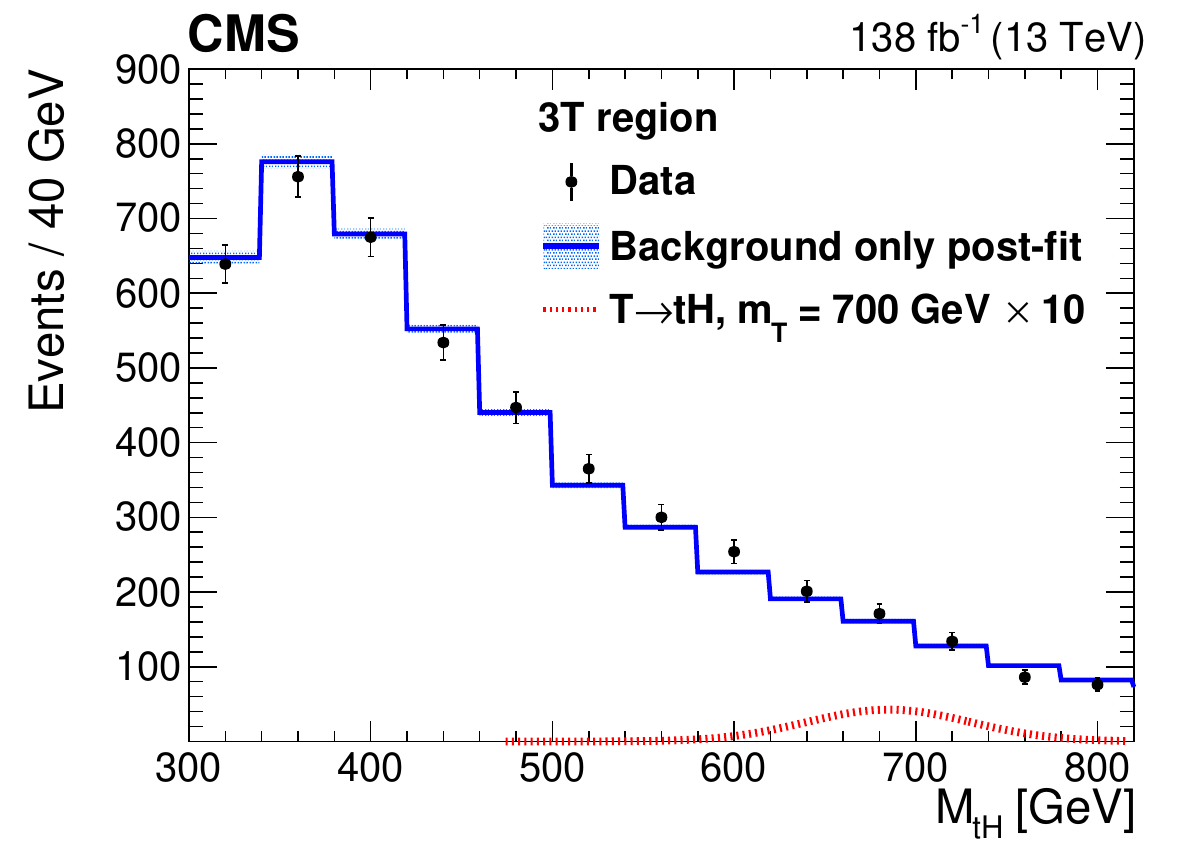}
    \includegraphics[width=0.49\textwidth]{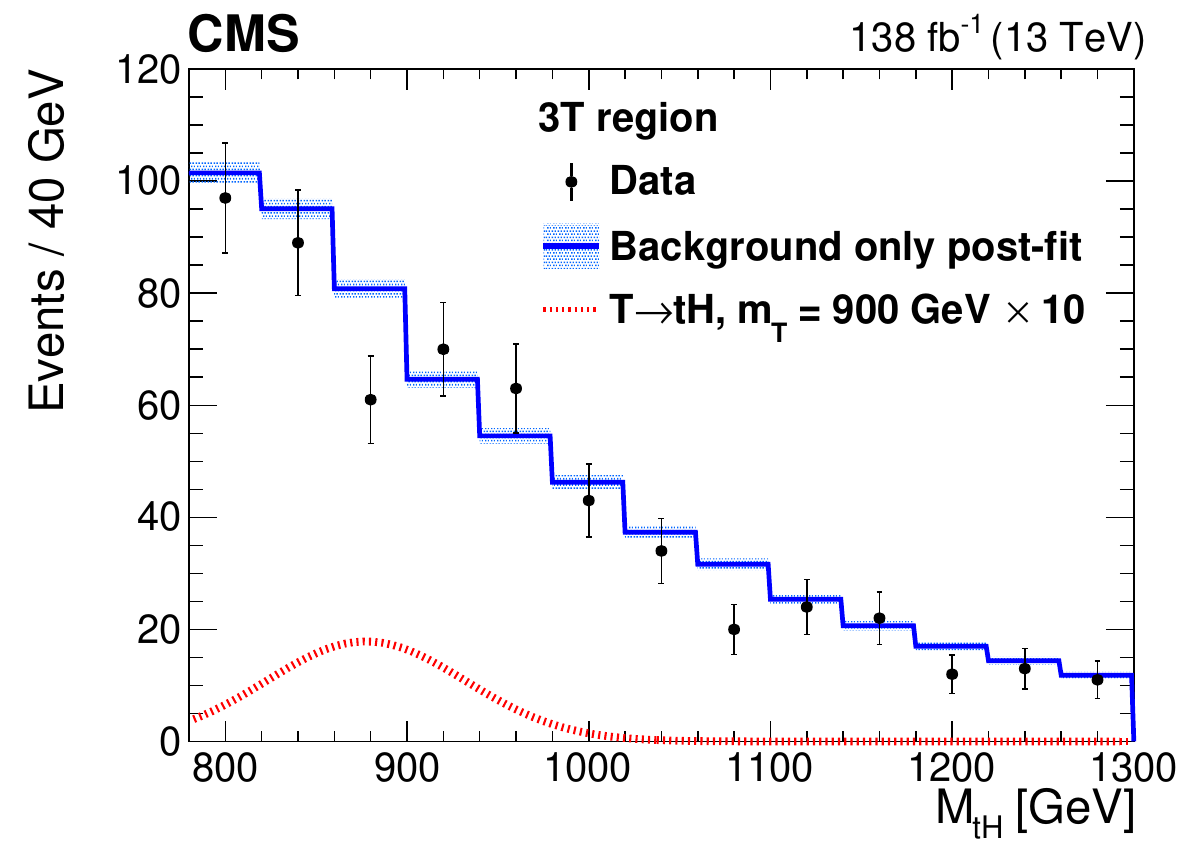}
    \caption{
        Five-jet invariant mass distributions after a background-only fit (blue histogram) to the complete dataset (black markers) in the 2M1L (upper), 3M (middle), and 3T (lower) regions for low-mass (left) and the high-mass (right) selections.
        The dashed blue band represents the uncertainty on the fitted background estimate, and red dashed line shows the expected signal distribution for a 700\GeV (low-mass selection) and a 900\GeV (high-mass selection) \tprime quark.
        The fit is performed on the combined data from all three years in the all-\tH channel.
    }
    \label{fig:fh-PostFit1}
\end{figure*}

\begin{figure}[!hbtp]
    \centering
    \includegraphics[width=0.49\textwidth]{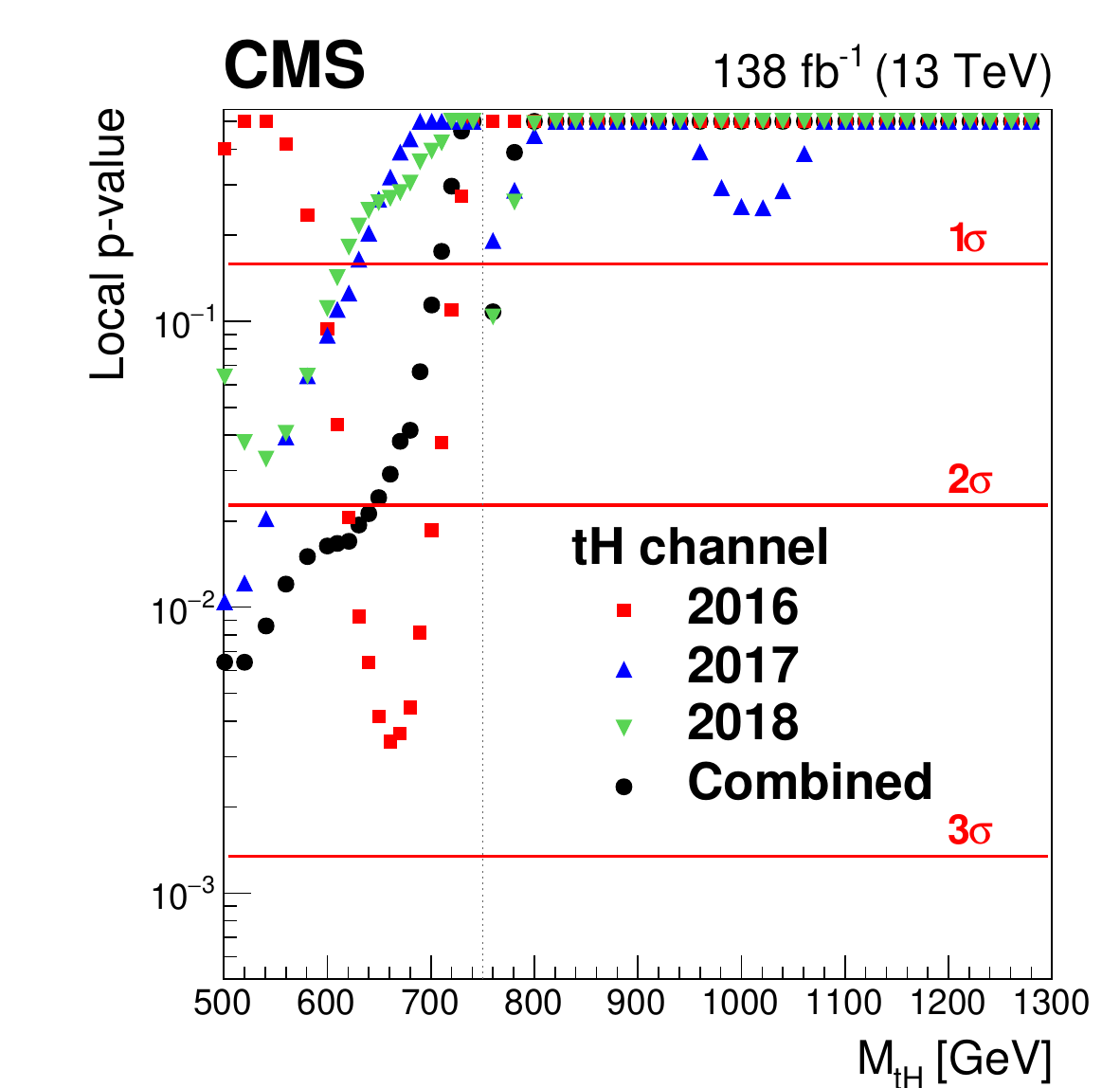}
    \caption{
        Observed $p$-values when considering the \tH channel for each year and their combination.
    }
    \label{fig:PValue}
\end{figure}

Figure~\ref{fig:fh-PostFit1} displays the background-only fits in each of the \tH channel search regions for the combined 2016, 2017, and 2018 data and both the low-mass and high-mass selections.
Each figure includes the expected signal distribution for a 700\GeV (low-mass selection) or 900\GeV (high-mass selection) \tprime quark.
An upward fluctuation near 680\GeV of the \tH spectrum is observed in 2016
data that vanishes once all data-taking years are combined.
Figure~\ref{fig:PValue} presents the observed $p$-values~\cite{pvalue}.

\begin{figure*}[!htb]
    \centering
    \includegraphics[width=0.49\textwidth]{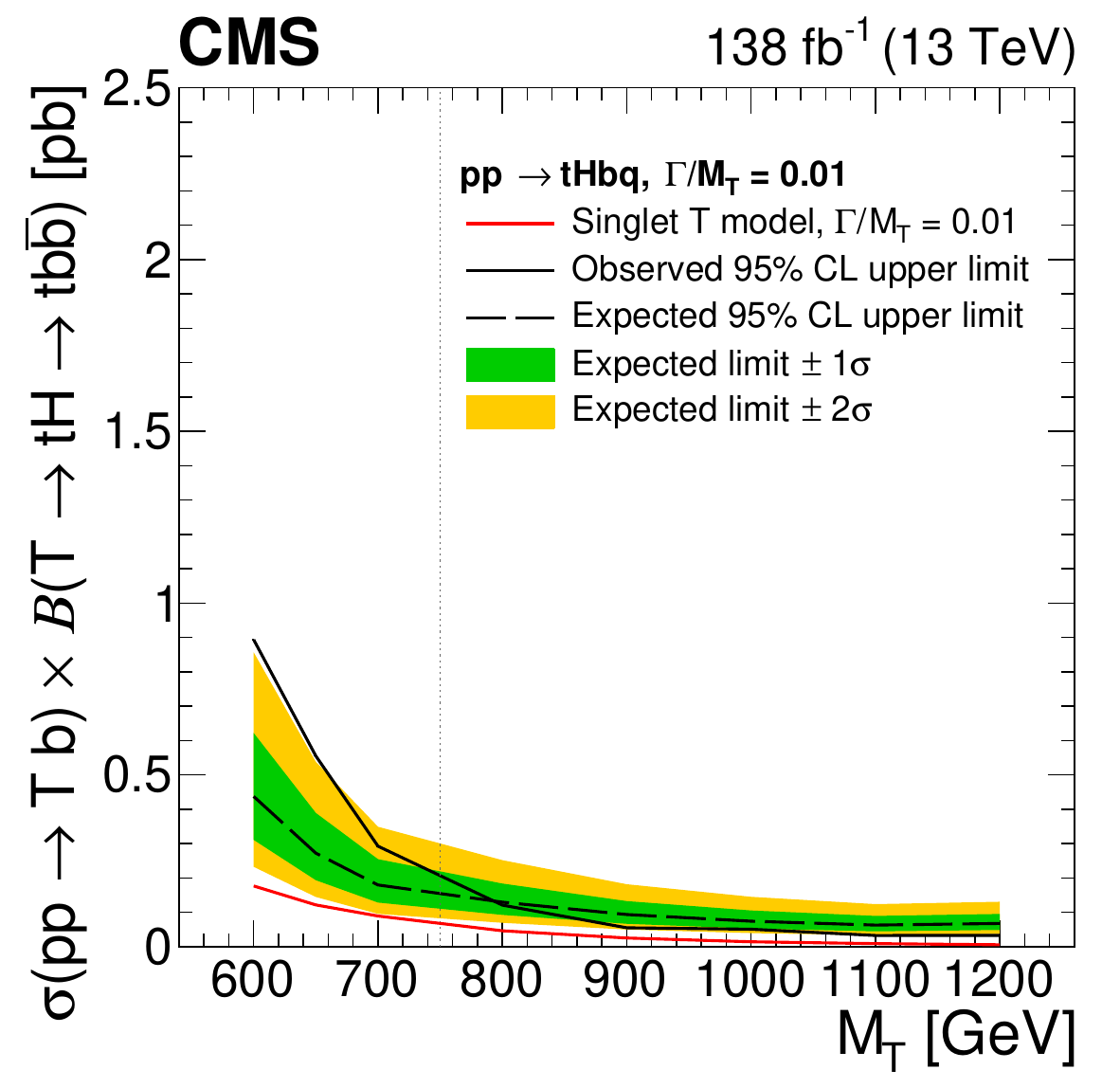}
    \includegraphics[width=0.49\textwidth]{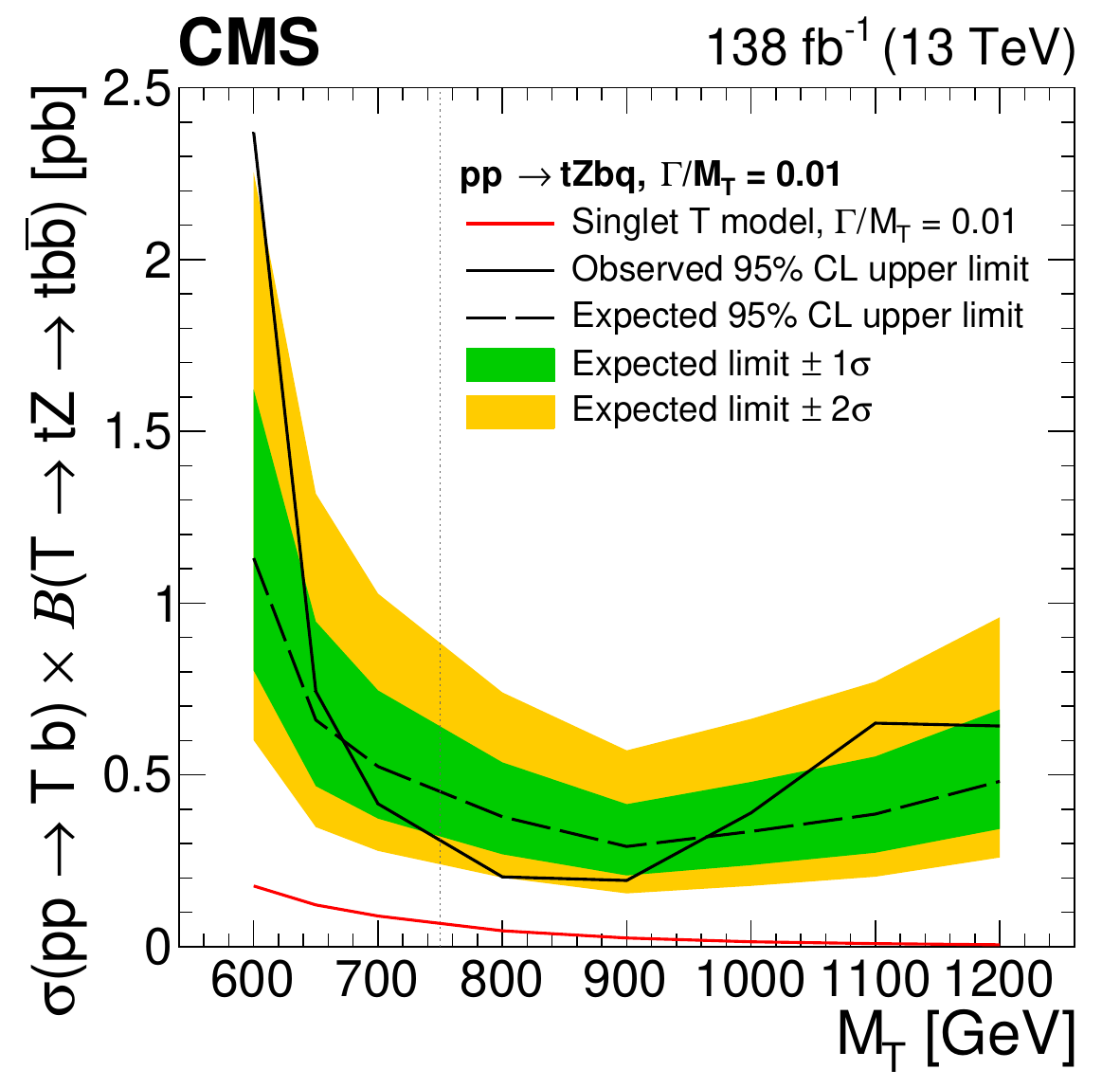}
    \includegraphics[width=0.49\textwidth]{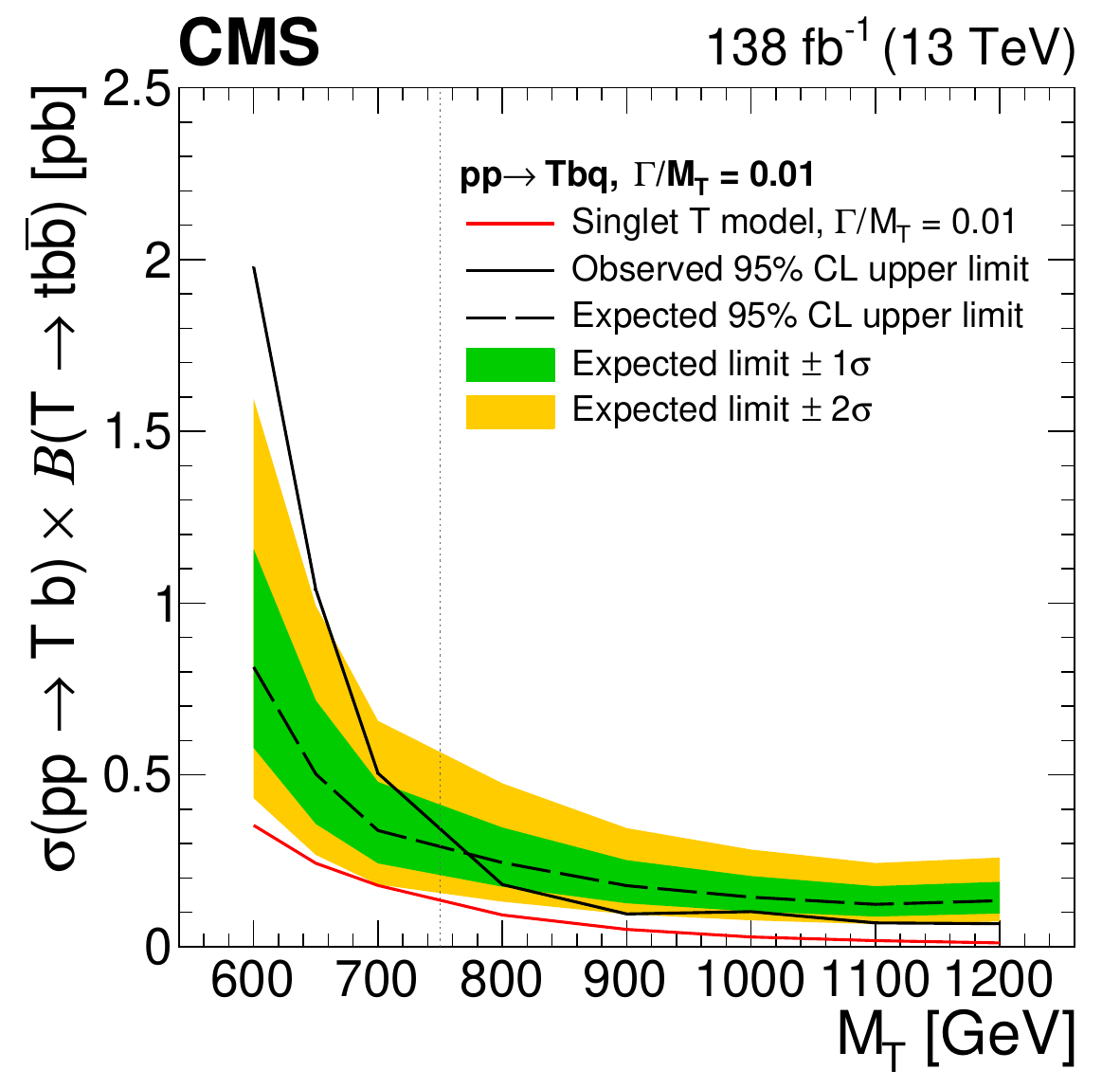}
    \includegraphics[width=0.49\textwidth]{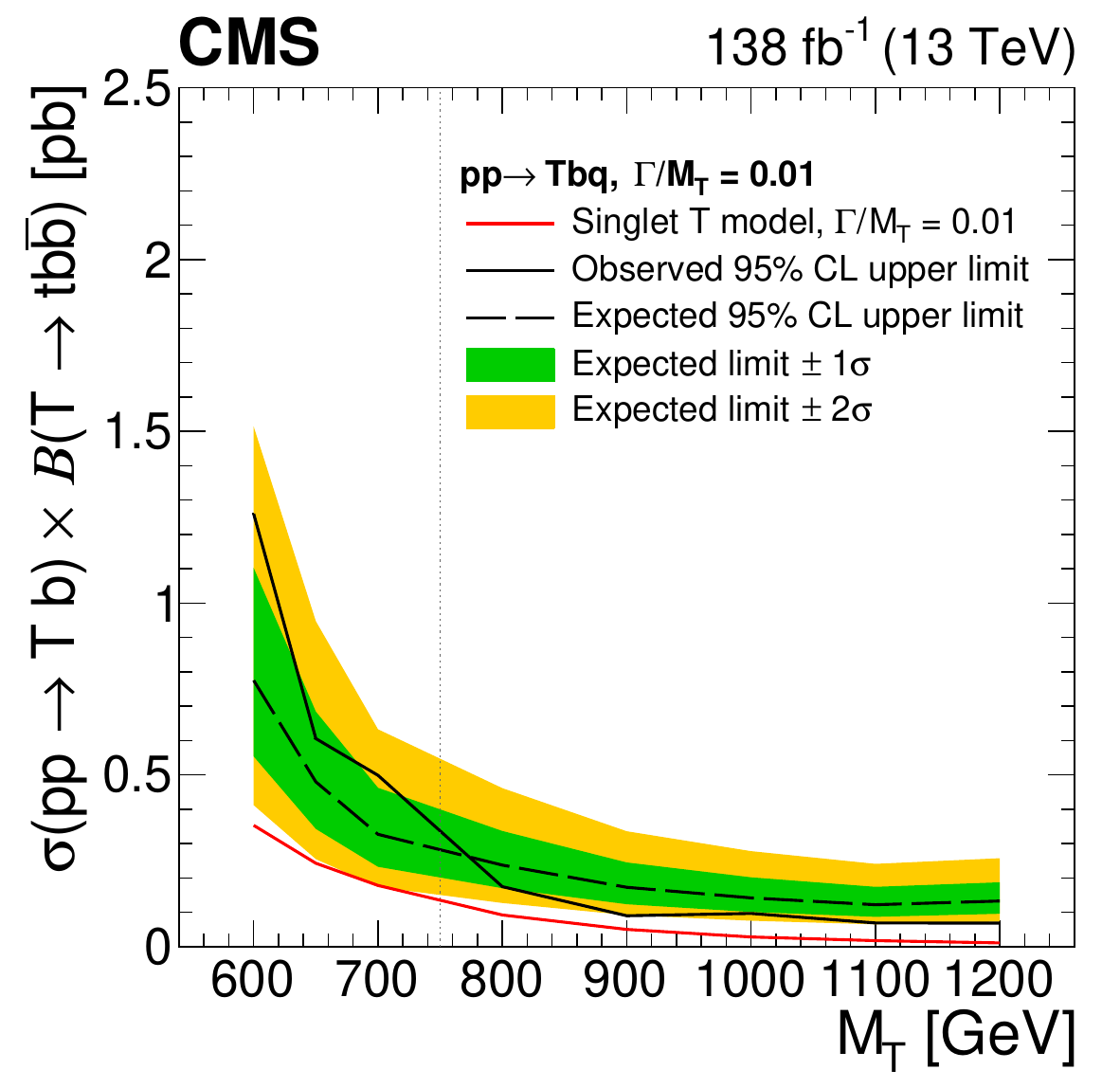}
    \caption{
        The observed and expected 95\% \CL limits on the cross section for associated production with a \PQb quark for final states \tHbq (upper left), \tZbq (upper right), their sum $\tHbq$+$\tZbq$ (lower left), and $\tHbq$+$\tZbq$ including the leakage of \tH events into the \tZ channel (lower right) for different assumed values of the \tprime quark mass.
        The vertical dashed line represents the crossover point in sensitivity: for masses to the left, the low-mass selection is used to set limits, while for masses to the right, the high-mass selection is used to set limits.
        The red lines indicate the theoretical cross section for the singlet model~\cite{AguilarSaavedra:2009es}.
    }
    \label{fig:massLim_Tbq1}
\end{figure*}

To evaluate the 95\% confidence level (\CL) observed and expected limits on the cross section for \tprime quark production, we follow the LHC \CLs criterion~\cite{CLS2,CLS1} using the profile likelihood ratio test statistic~\cite{LHCCLs} and the asymptotic formula~\cite{AsympCLs}.
The 95\% \CL upper limits on the cross sections found from the search signatures are derived for \tprime quark masses from 600 to 1200\GeV.
The limits are computed for each of the channels: \tH, \tZ, their combination, and the combination of all-\tH channel and \tZ channel, as shown in Fig.~\ref{fig:massLim_Tbq1}.

For \tprime quark masses above 700\GeV, the observed limits are consistent with the expected ones.
Across the full considered \tprime quark mass range of $[600, 1200]\GeV$, the narrow-width approximation model cannot be excluded with the current data set.
Tabulated results are provided in the HEPData record for this analysis~\cite{hepdata}.
\section{Summary}
\label{sec:summary}

A search for a vector-like top quark \tprime in the single production mode was performed using proton-proton collision events at $\sqrt{s} = 13\TeV$ collected by the CMS experiment in 2016--2018.
In this search, the \tprime quark is assumed to couple only to standard model third-generation quarks.
We consider signatures containing a top quark and a Higgs ($\tH$) or \PZ ($\tZ$) boson decaying to a bottom quark-antiquark pair.
The major background processes are top quark-antiquark pair and multijet production.
The feature in the \tH final state found in the previous search~\cite{Sirunyan:2019xeh} is not confirmed with a larger dataset and improved event selection.
No evidence for the \tprime quark production in the $\Tbq$ process is seen and 95\% confidence level upper limits are set on the product of the production cross section and branching fraction to \tH and \tZ that range from 1260 to 68\unit{fb} for \tprime quark masses of 600--1200\GeV.
The limits are stronger than those in the previous search by at least a factor of three.

\begin{acknowledgments}
    We congratulate our colleagues in the CERN accelerator departments for the excellent performance of the LHC and thank the technical and administrative staffs at CERN and at other CMS institutes for their contributions to the success of the CMS effort. In addition, we gratefully acknowledge the computing centers and personnel of the Worldwide LHC Computing Grid and other centers for delivering so effectively the computing infrastructure essential to our analyses. Finally, we acknowledge the enduring support for the construction and operation of the LHC, the CMS detector, and the supporting computing infrastructure provided by the following funding agencies: SC (Armenia), BMBWF and FWF (Austria); FNRS and FWO (Belgium); CNPq, CAPES, FAPERJ, FAPERGS, and FAPESP (Brazil); MES and BNSF (Bulgaria); CERN; CAS, MoST, and NSFC (China); MINCIENCIAS (Colombia); MSES and CSF (Croatia); RIF (Cyprus); SENESCYT (Ecuador); ERC PRG, RVTT3 and MoER TK202 (Estonia); Academy of Finland, MEC, and HIP (Finland); CEA and CNRS/IN2P3 (France); SRNSF (Georgia); BMBF, DFG, and HGF (Germany); GSRI (Greece); NKFIH (Hungary); DAE and DST (India); IPM (Iran); SFI (Ireland); INFN (Italy); MSIP and NRF (Republic of Korea); MES (Latvia); LMTLT (Lithuania); MOE and UM (Malaysia); BUAP, CINVESTAV, CONACYT, LNS, SEP, and UASLP-FAI (Mexico); MOS (Montenegro); MBIE (New Zealand); PAEC (Pakistan); MES and NSC (Poland); FCT (Portugal); MESTD (Serbia); MCIN/AEI and PCTI (Spain); MOSTR (Sri Lanka); Swiss Funding Agencies (Switzerland); MST (Taipei); MHESI and NSTDA (Thailand); TUBITAK and TENMAK (Turkey); NASU (Ukraine); STFC (United Kingdom); DOE and NSF (USA).

    \hyphenation{Rachada-pisek} Individuals have received support from the Marie-Curie program and the European Research Council and Horizon 2020 Grant, contract Nos.\ 675440, 724704, 752730, 758316, 765710, 824093, 101115353, 101002207, and COST Action CA16108 (European Union); the Leventis Foundation; the Alfred P.\ Sloan Foundation; the Alexander von Humboldt Foundation; the Science Committee, project no. 22rl-037 (Armenia); the Belgian Federal Science Policy Office; the Fonds pour la Formation \`a la Recherche dans l'Industrie et dans l'Agriculture (FRIA-Belgium); the Agentschap voor Innovatie door Wetenschap en Technologie (IWT-Belgium); the F.R.S.-FNRS and FWO (Belgium) under the ``Excellence of Science -- EOS" -- be.h project n.\ 30820817; the Beijing Municipal Science \& Technology Commission, No. Z191100007219010 and Fundamental Research Funds for the Central Universities (China); the Ministry of Education, Youth and Sports (MEYS) of the Czech Republic; the Shota Rustaveli National Science Foundation, grant FR-22-985 (Georgia); the Deutsche Forschungsgemeinschaft (DFG), under Germany's Excellence Strategy -- EXC 2121 ``Quantum Universe" -- 390833306, and under project number 400140256 - GRK2497; the Hellenic Foundation for Research and Innovation (HFRI), Project Number 2288 (Greece); the Hungarian Academy of Sciences, the New National Excellence Program - \'UNKP, the NKFIH research grants K 131991, K 133046, K 138136, K 143460, K 143477, K 146913, K 146914, K 147048, 2020-2.2.1-ED-2021-00181, and TKP2021-NKTA-64 (Hungary); the Council of Science and Industrial Research, India; ICSC -- National Research Center for High Performance Computing, Big Data and Quantum Computing and FAIR -- Future Artificial Intelligence Research, funded by the NextGenerationEU program (Italy); the Latvian Council of Science; the Ministry of Education and Science, project no. 2022/WK/14, and the National Science Center, contracts Opus 2021/41/B/ST2/01369 and 2021/43/B/ST2/01552 (Poland); the Funda\c{c}\~ao para a Ci\^encia e a Tecnologia, grant CEECIND/01334/2018 (Portugal); the National Priorities Research Program by Qatar National Research Fund; MCIN/AEI/10.13039/501100011033, ERDF ``a way of making Europe", and the Programa Estatal de Fomento de la Investigaci{\'o}n Cient{\'i}fica y T{\'e}cnica de Excelencia Mar\'{\i}a de Maeztu, grant MDM-2017-0765 and Programa Severo Ochoa del Principado de Asturias (Spain); the Chulalongkorn Academic into Its 2nd Century Project Advancement Project, and the National Science, Research and Innovation Fund via the Program Management Unit for Human Resources \& Institutional Development, Research and Innovation, grant B37G660013 (Thailand); the Kavli Foundation; the Nvidia Corporation; the SuperMicro Corporation; the Welch Foundation, contract C-1845; and the Weston Havens Foundation (USA).
\end{acknowledgments}

\bibliography{auto_generated}

\providecommand{\href}[2]{#2}\begingroup\raggedright\begin{thebibliography}{10}%
\makeatletter
\providecommand{\hrefCMSnoop }[0]{\@secondoftwo}%
\makeatother
\providecommand{\doi}{\texttt{doi:}\begingroup \urlstyle{tt}\Url}

\bibitem{Okada:2012gy}
\hrefCMSnoop {}{Y.~Okada and L.~Panizzi, ``{LHC signatures of vector-like
  quarks}'',} \textit{ Adv. High Energy Phys.} \textbf{ 2013} (2013) 364936,
  \href{http://dx.doi.org/10.1155/2013/364936}{\doi{10.1155/2013/364936}},
  \href{http://www.arXiv.org/abs/1207.5607}{\texttt{arXiv:1207.5607}}.

\bibitem{Aguilar-Saavedra:2013qpa}
\hrefCMSnoop {}{J.~A. Aguilar-Saavedra, R.~Benbrik, S.~Heinemeyer, and
  M.~P{\'e}rez-Victoria, ``{Handbook of vectorlike quarks: Mixing and single
  production}'',} \textit{ Phys. Rev. D} \textbf{ 88} (2013) 094010,
  \href{http://dx.doi.org/10.1103/PhysRevD.88.094010}{\doi{10.1103/PhysRevD.88.094010}},
  \href{http://www.arXiv.org/abs/1306.0572}{\texttt{arXiv:1306.0572}}.

\bibitem{DeSimone:2012fs}
\hrefCMSnoop {}{A.~De~Simone, O.~Matsedonskyi, R.~Rattazzi, and A.~Wulzer, ``A
  first top partner hunter's guide'',} \textit{ JHEP} \textbf{ 04} (2013) 004,
  \href{http://dx.doi.org/10.1007/JHEP04(2013)004}{\doi{10.1007/JHEP04(2013)004}},
  \href{http://www.arXiv.org/abs/1211.5663}{\texttt{arXiv:1211.5663}}.

\bibitem{Buchkremer:2013bha}
\hrefCMSnoop {}{M.~Buchkremer, G.~Cacciapaglia, A.~Deandrea, and L.~Panizzi,
  ``Model independent framework for searches of top partners'',} \textit{ Nucl.
  Phys. B} \textbf{ 876} (2013) 376,
  \href{http://dx.doi.org/10.1016/j.nuclphysb.2013.08.010}{\doi{10.1016/j.nuclphysb.2013.08.010}},
  \href{http://www.arXiv.org/abs/1305.4172}{\texttt{arXiv:1305.4172}}.

\bibitem{Djouadi:2012ae}
\hrefCMSnoop {}{A.~Djouadi and A.~Lenz, ``{Sealing the fate of a fourth
  generation of fermions}'',} \textit{ Phys. Lett. B} \textbf{ 715} (2012) 310,
  \href{http://dx.doi.org/10.1016/j.physletb.2012.07.060}{\doi{10.1016/j.physletb.2012.07.060}},
  \href{http://www.arXiv.org/abs/1204.1252}{\texttt{arXiv:1204.1252}}.

\bibitem{Lenz:2013iha}
\hrefCMSnoop {}{A.~Lenz, ``{Constraints on a fourth generation of fermions from
  Higgs Boson searches}'',} \textit{ Adv. High Energy Phys.} \textbf{ 2013}
  (2013) 910275,
  \href{http://dx.doi.org/10.1155/2013/910275}{\doi{10.1155/2013/910275}}.

\bibitem{Chatrchyan:2012fp}
\hrefCMSnoop {}{{CMS Collaboration}, ``{Combined search for the quarks of a
  sequential fourth generation}'',} \textit{ Phys. Rev. D} \textbf{ 86} (2012)
  112003,
  \href{http://dx.doi.org/10.1103/PhysRevD.86.112003}{\doi{10.1103/PhysRevD.86.112003}},
  \href{http://www.arXiv.org/abs/1209.1062}{\texttt{arXiv:1209.1062}}.

\bibitem{Vignaroli:2012nf}
\hrefCMSnoop {}{N.~Vignaroli, ``{Early discovery of top partners and test of
  the Higgs nature}'',} \textit{ Phys. Rev. D} \textbf{ 86} (2012) 075017,
  \href{http://dx.doi.org/10.1103/PhysRevD.86.075017}{\doi{10.1103/PhysRevD.86.075017}},
  \href{http://www.arXiv.org/abs/1207.0830}{\texttt{arXiv:1207.0830}}.

\bibitem{AguilarSaavedra:2009es}
\hrefCMSnoop {}{J.~A. Aguilar-Saavedra, ``{Identifying top partners at LHC}'',}
  \textit{ JHEP} \textbf{ 11} (2009) 030,
  \href{http://dx.doi.org/10.1088/1126-6708/2009/11/030}{\doi{10.1088/1126-6708/2009/11/030}},
  \href{http://www.arXiv.org/abs/0907.3155}{\texttt{arXiv:0907.3155}}.

\bibitem{Sirunyan:2017pks}
\hrefCMSnoop {}{{CMS Collaboration}, ``{Search for pair production of
  vector-like quarks in the $\Pb\PW\PAQb\PW$ channel from proton-proton
  collisions at $\sqrt{s} = 13\TeV$}'',} \textit{ Phys. Lett. B} \textbf{ 779}
  (2018) 82,
  \href{http://dx.doi.org/10.1016/j.physletb.2018.01.077}{\doi{10.1016/j.physletb.2018.01.077}},
  \href{http://www.arXiv.org/abs/1710.01539}{\texttt{arXiv:1710.01539}}.

\bibitem{ATLAS:2018uky}
\hrefCMSnoop {}{{ATLAS Collaboration}, ``{Search for pair production of heavy
  vector-like quarks decaying into hadronic final states in $pp$ collisions at
  $\sqrt{s} = 13$ TeV with the ATLAS detector}'',} \textit{ Phys. Rev. D}
  \textbf{ 98} (2018) 092005,
  \href{http://dx.doi.org/10.1103/PhysRevD.98.092005}{\doi{10.1103/PhysRevD.98.092005}},
  \href{http://www.arXiv.org/abs/1808.01771}{\texttt{arXiv:1808.01771}}.

\bibitem{ATLAS:2022tla}
\hrefCMSnoop {}{{ATLAS Collaboration}, ``{Search for pair-produced vector-like
  top and bottom partners in events with large missing transverse momentum in
  pp collisions with the ATLAS detector}'',} \textit{ Eur. Phys. J. C} \textbf{
  83} (2023) 719,
  \href{http://dx.doi.org/10.1140/epjc/s10052-023-11790-7}{\doi{10.1140/epjc/s10052-023-11790-7}},
  \href{http://www.arXiv.org/abs/2212.05263}{\texttt{arXiv:2212.05263}}.

\bibitem{CMS:2022fck}
\hrefCMSnoop {}{{CMS Collaboration}, ``{Search for pair production of
  vector-like quarks in leptonic final states in proton-proton collisions at
  $\sqrt{s} = 13\TeV$}'',} \textit{ JHEP} \textbf{ 07} (2023) 020,
  \href{http://dx.doi.org/10.1007/JHEP07(2023)020}{\doi{10.1007/JHEP07(2023)020}},
  \href{http://www.arXiv.org/abs/2209.07327}{\texttt{arXiv:2209.07327}}.

\bibitem{ATLAS:2022hnn}
\hrefCMSnoop {}{{ATLAS Collaboration}, ``{Search for pair-production of
  vector-like quarks in $\Pp\Pp$ collision events at $\sqrt{s}=13\TeV$ with at
  least one leptonically decaying \PZ boson and a third-generation quark with
  the ATLAS detector}'',} \textit{ Phys. Lett. B} \textbf{ 843} (2023) 138019,
  \href{http://dx.doi.org/10.1016/j.physletb.2023.138019}{\doi{10.1016/j.physletb.2023.138019}},
  \href{http://www.arXiv.org/abs/2210.15413}{\texttt{arXiv:2210.15413}}.

\bibitem{ATLAS:2018dyh}
\hrefCMSnoop {}{{ATLAS Collaboration}, ``{Search for single production of
  vector-like quarks decaying into $Wb$ in $pp$ collisions at $\sqrt{s} = 13$
  TeV with the ATLAS detector}'',} \textit{ JHEP} \textbf{ 05} (2019) 164,
  \href{http://dx.doi.org/10.1007/JHEP05(2019)164}{\doi{10.1007/JHEP05(2019)164}},
  \href{http://www.arXiv.org/abs/1812.07343}{\texttt{arXiv:1812.07343}}.

\bibitem{ATLAS:2018cjd}
\hrefCMSnoop {}{{ATLAS Collaboration}, ``{Search for large missing transverse
  momentum in association with one top-quark in proton-proton collisions at $
  \sqrt{s} $ = 13 TeV with the ATLAS detector}'',} \textit{ JHEP} \textbf{ 05}
  (2019) 041,
  \href{http://dx.doi.org/10.1007/JHEP05(2019)041}{\doi{10.1007/JHEP05(2019)041}},
  \href{http://www.arXiv.org/abs/1812.09743}{\texttt{arXiv:1812.09743}}.

\bibitem{ATLAS:2022ozf}
\hrefCMSnoop {}{{ATLAS Collaboration}, ``{Search for single production of a
  vectorlike $T$ quark decaying into a Higgs boson and top quark with fully
  hadronic final states using the ATLAS detector}'',} \textit{ Phys. Rev. D}
  \textbf{ 105} (2022) 092012,
  \href{http://dx.doi.org/10.1103/PhysRevD.105.092012}{\doi{10.1103/PhysRevD.105.092012}},
  \href{http://www.arXiv.org/abs/2201.07045}{\texttt{arXiv:2201.07045}}.

\bibitem{Sirunyan:2017tfc}
\hrefCMSnoop {}{{CMS Collaboration}, ``{Search for single production of
  vector-like quarks decaying into a \PQb quark and a \PW boson in
  proton-proton collisions at $\sqrt{s} = 13\TeV$}'',} \textit{ Phys. Lett. B}
  \textbf{ 772} (2017) 634,
  \href{http://dx.doi.org/10.1016/j.physletb.2017.07.022}{\doi{10.1016/j.physletb.2017.07.022}},
  \href{http://www.arXiv.org/abs/1701.08328}{\texttt{arXiv:1701.08328}}.

\bibitem{Sirunyan:2017ynj}
\hrefCMSnoop {}{{CMS Collaboration}, ``{Search for single production of a
  vector-like \tprime quark decaying to a \PZ boson and a top quark in
  proton-proton collisions at $\sqrt{s} = 13\TeV$}'',} \textit{ Phys. Lett. B}
  \textbf{ 781} (2018) 574,
  \href{http://dx.doi.org/10.1016/j.physletb.2018.04.036}{\doi{10.1016/j.physletb.2018.04.036}},
  \href{http://www.arXiv.org/abs/1708.01062}{\texttt{arXiv:1708.01062}}.

\bibitem{CMS:2022yxp}
\hrefCMSnoop {}{{CMS Collaboration}, ``{Search for single production of a
  vector-like T quark decaying to a top quark and a \PZ boson in the final
  state with jets and missing transverse momentum at $\sqrt{s} = 13\TeV$}'',}
  \textit{ JHEP} \textbf{ 05} (2022) 093,
  \href{http://dx.doi.org/10.1007/JHEP05(2022)093}{\doi{10.1007/JHEP05(2022)093}},
  \href{http://www.arXiv.org/abs/2201.02227}{\texttt{arXiv:2201.02227}}.

\bibitem{Khachatryan:2016vph}
\hrefCMSnoop {}{{CMS Collaboration}, ``{Search for single production of a heavy
  vector-like \tprime quark decaying to a Higgs boson and a top quark with a
  lepton and jets in the final state}'',} \textit{ Phys. Lett. B} \textbf{ 771}
  (2017) 80,
  \href{http://dx.doi.org/10.1016/j.physletb.2017.05.019}{\doi{10.1016/j.physletb.2017.05.019}},
  \href{http://www.arXiv.org/abs/1612.00999}{\texttt{arXiv:1612.00999}}.

\bibitem{Sirunyan:2016ipo}
\hrefCMSnoop {}{{CMS Collaboration}, ``{Search for electroweak production of a
  vector-like quark decaying to a top quark and a Higgs boson using boosted
  topologies in fully hadronic final states}'',} \textit{ JHEP} \textbf{ 04}
  (2017) 136,
  \href{http://dx.doi.org/10.1007/JHEP04(2017)136}{\doi{10.1007/JHEP04(2017)136}},
  \href{http://www.arXiv.org/abs/1612.05336}{\texttt{arXiv:1612.05336}}.

\bibitem{Sirunyan:2019xeh}
\hrefCMSnoop {}{{CMS Collaboration}, ``{Search for electroweak production of a
  vector-like \tprime quark using fully hadronic final states}'',} \textit{
  JHEP} \textbf{ 01} (2020) 036,
  \href{http://dx.doi.org/10.1007/JHEP01(2020)036}{\doi{10.1007/JHEP01(2020)036}},
  \href{http://www.arXiv.org/abs/1909.04721}{\texttt{arXiv:1909.04721}}.

\bibitem{CMS:2014pgm}
\hrefCMSnoop {}{{CMS Collaboration}, ``{Description and performance of track
  and primary-vertex reconstruction with the CMS tracker}'',} \textit{ JINST}
  \textbf{ 9} (2014) P10009,
  \href{http://dx.doi.org/10.1088/1748-0221/9/10/P10009}{\doi{10.1088/1748-0221/9/10/P10009}},
  \href{http://www.arXiv.org/abs/1405.6569}{\texttt{arXiv:1405.6569}}.

\bibitem{Phase1Pixel}
\hrefCMSnoop {}{{CMS Tracker Group} Collaboration, ``The {CMS} phase-1 pixel
  detector upgrade'',} \textit{ JINST} \textbf{ 16} (2021) P02027,
  \href{http://dx.doi.org/10.1088/1748-0221/16/02/P02027}{\doi{10.1088/1748-0221/16/02/P02027}},
  \href{http://www.arXiv.org/abs/2012.14304}{\texttt{arXiv:2012.14304}}.

\bibitem{DP-2020-049}
\href {https://cds.cern.ch/record/2743740}{{CMS Collaboration}, ``Track impact
  parameter resolution for the full pseudo rapidity coverage in the 2017
  dataset with the {CMS} phase-1 pixel detector'',} CMS Detector Performance
  Summary CMS-DP-2020-049, 2020.

\bibitem{CMS:2020uim}
\hrefCMSnoop {}{{CMS Collaboration}, ``Electron and photon reconstruction and
  identification with the {CMS} experiment at the {CERN} {LHC}'',} \textit{
  JINST} \textbf{ 16} (2021) P05014,
  \href{http://dx.doi.org/10.1088/1748-0221/16/05/P05014}{\doi{10.1088/1748-0221/16/05/P05014}},
  \href{http://www.arXiv.org/abs/2012.06888}{\texttt{arXiv:2012.06888}}.

\bibitem{CMS-DP-2020-021}
\href {https://cds.cern.ch/record/2717925}{{CMS Collaboration}, ``{ECAL} 2016
  refined calibration and {Run 2} summary plots'',} CMS Detector Performance
  Summary CMS-DP-2020-021, 2020.

\bibitem{Chatrchyan:2008zzk}
\hrefCMSnoop {}{{CMS Collaboration}, ``The {CMS} experiment at the {CERN}
  {LHC}'',} \textit{ JINST} \textbf{ 3} (2008) S08004,
  \href{http://dx.doi.org/10.1088/1748-0221/3/08/S08004}{\doi{10.1088/1748-0221/3/08/S08004}}.

\bibitem{Khachatryan:2016bia}
\hrefCMSnoop {}{{CMS Collaboration}, ``{The CMS trigger system}'',} \textit{
  JINST} \textbf{ 12} (2017) P01020,
  \href{http://dx.doi.org/10.1088/1748-0221/12/01/P01020}{\doi{10.1088/1748-0221/12/01/P01020}},
  \href{http://www.arXiv.org/abs/1609.02366}{\texttt{arXiv:1609.02366}}.

\bibitem{CMS-TDR-15-02}
\href {http://cds.cern.ch/record/2020886}{{CMS Collaboration}, ``Technical
  proposal for the {Phase-II} upgrade of the {Compact Muon Solenoid}'',} CMS
  Technical Proposal CERN-LHCC-2015-010, CMS-TDR-15-02, 2015.

\bibitem{Sirunyan:2017ulk}
\hrefCMSnoop {}{{CMS Collaboration}, ``{Particle-flow reconstruction and global
  event description with the CMS detector}'',} \textit{ JINST} \textbf{ 12}
  (2017) P10003,
  \href{http://dx.doi.org/10.1088/1748-0221/12/10/P10003}{\doi{10.1088/1748-0221/12/10/P10003}},
  \href{http://www.arXiv.org/abs/1706.04965}{\texttt{arXiv:1706.04965}}.

\bibitem{Cacciari:2008gp}
\hrefCMSnoop {}{M.~Cacciari, G.~P. Salam, and G.~Soyez, ``The anti-$\kt$ jet
  clustering algorithm'',} \textit{ JHEP} \textbf{ 04} (2008) 063,
  \href{http://dx.doi.org/10.1088/1126-6708/2008/04/063}{\doi{10.1088/1126-6708/2008/04/063}},
  \href{http://www.arXiv.org/abs/0802.1189}{\texttt{arXiv:0802.1189}}.

\bibitem{Cacciari:2011ma}
\hrefCMSnoop {}{M.~Cacciari, G.~P. Salam, and G.~Soyez, ``{FastJet user
  manual}'',} \textit{ Eur. Phys. J. C} \textbf{ 72} (2012) 1896,
  \href{http://dx.doi.org/10.1140/epjc/s10052-012-1896-2}{\doi{10.1140/epjc/s10052-012-1896-2}},
  \href{http://www.arXiv.org/abs/1111.6097}{\texttt{arXiv:1111.6097}}.

\bibitem{CMS:2020ebo}
\hrefCMSnoop {}{{CMS Collaboration}, ``{Pileup mitigation at CMS in 13\TeV
  data}'',} \textit{ JINST} \textbf{ 15} (2020) P09018,
  \href{http://dx.doi.org/10.1088/1748-0221/15/09/P09018}{\doi{10.1088/1748-0221/15/09/P09018}},
  \href{http://www.arXiv.org/abs/2003.00503}{\texttt{arXiv:2003.00503}}.

\bibitem{CMS:2016lmd}
\hrefCMSnoop {}{{CMS Collaboration}, ``{Jet energy scale and resolution in the
  CMS experiment in $\Pp\Pp$ collisions at 8\TeV}'',} \textit{ JINST} \textbf{
  12} (2017) P02014,
  \href{http://dx.doi.org/10.1088/1748-0221/12/02/P02014}{\doi{10.1088/1748-0221/12/02/P02014}},
  \href{http://www.arXiv.org/abs/1607.03663}{\texttt{arXiv:1607.03663}}.

\bibitem{BTV-16-002}
\hrefCMSnoop {}{{CMS Collaboration}, ``{Identification of heavy-flavour jets
  with the CMS detector in $\Pp\Pp$ collisions at 13\TeV}'',} \textit{ JINST}
  \textbf{ 13} (2018) P05011,
  \href{http://dx.doi.org/10.1088/1748-0221/13/05/P05011}{\doi{10.1088/1748-0221/13/05/P05011}},
  \href{http://www.arXiv.org/abs/1712.07158}{\texttt{arXiv:1712.07158}}.

\bibitem{MG5_aMCNLO}
J.~Alwall\hrefCMSnoop {}{ { et~al.}, ``{The automated computation of tree-level
  and next-to-leading order differential cross sections, and their matching to
  parton shower simulations}'',} \textit{ JHEP} \textbf{ 07} (2014) 079,
  \href{http://dx.doi.org/10.1007/JHEP07(2014)079}{\doi{10.1007/JHEP07(2014)079}},
  \href{http://www.arXiv.org/abs/1405.0301}{\texttt{arXiv:1405.0301}}.

\bibitem{PDG2022}
\hrefCMSnoop {}{{Particle Data Group} Collaboration, ``Review of particle
  physics'',} \textit{ Prog. Theor. Exp. Phys.} \textbf{ 2022} (2022) 083C01,
  \href{http://dx.doi.org/10.1093/ptep/ptac097}{\doi{10.1093/ptep/ptac097}}.

\bibitem{Sjostrand:2014zea}
T.~Sj{\"o}strand\hrefCMSnoop {}{ { et~al.}, ``{An Introduction to
  {\PYTHIA}8.2}'',} \textit{ Comput. Phys. Commun.} \textbf{ 191} (2015) 159,
  \href{http://dx.doi.org/10.1016/j.cpc.2015.01.024}{\doi{10.1016/j.cpc.2015.01.024}},
  \href{http://www.arXiv.org/abs/1410.3012}{\texttt{arXiv:1410.3012}}.

\bibitem{Khachatryan:2015pea}
\hrefCMSnoop {}{{CMS Collaboration}, ``{Event generator tunes obtained from
  underlying event and multiparton scattering measurements}'',} \textit{ Eur.
  Phys. J. C} \textbf{ 76} (2016) 155,
  \href{http://dx.doi.org/10.1140/epjc/s10052-016-3988-x}{\doi{10.1140/epjc/s10052-016-3988-x}},
  \href{http://www.arXiv.org/abs/1512.00815}{\texttt{arXiv:1512.00815}}.

\bibitem{CMS-GEN-17-001}
\hrefCMSnoop {}{{CMS Collaboration}, ``{Extraction and validation of a new set
  of CMS PYTHIA8 tunes from underlying-event measurements}'',} \textit{ Eur.
  Phys. J. C} \textbf{ 80} (2020) 4,
  \href{http://dx.doi.org/10.1140/epjc/s10052-019-7499-4}{\doi{10.1140/epjc/s10052-019-7499-4}},
  \href{http://www.arXiv.org/abs/1903.12179}{\texttt{arXiv:1903.12179}}.

\bibitem{Ball:2014uwa}
\hrefCMSnoop {}{{NNPDF} Collaboration, ``{Parton distributions for the LHC Run
  II}'',} \textit{ JHEP} \textbf{ 04} (2015) 040,
  \href{http://dx.doi.org/10.1007/JHEP04(2015)040}{\doi{10.1007/JHEP04(2015)040}},
  \href{http://www.arXiv.org/abs/1410.8849}{\texttt{arXiv:1410.8849}}.

\bibitem{Ball:2017nwa}
\hrefCMSnoop {}{{NNPDF} Collaboration, ``{Parton distributions from
  high-precision collider data}'',} \textit{ Eur. Phys. J. C} \textbf{ 77}
  (2017) 663,
  \href{http://dx.doi.org/10.1140/epjc/s10052-017-5199-5}{\doi{10.1140/epjc/s10052-017-5199-5}},
  \href{http://www.arXiv.org/abs/1706.00428}{\texttt{arXiv:1706.00428}}.

\bibitem{Agostinelli:2002hh}
\hrefCMSnoop {}{{GEANT4} Collaboration, ``{\GEANTfour}---a simulation
  toolkit'',} \textit{ Nucl. Instrum. Meth. A} \textbf{ 506} (2003) 250,
  \href{http://dx.doi.org/10.1016/S0168-9002(03)01368-8}{\doi{10.1016/S0168-9002(03)01368-8}}.

\bibitem{CMS-LUM-17-003}
\hrefCMSnoop {}{{CMS Collaboration}, ``Precision luminosity measurement in
  proton-proton collisions at {$\sqrt{s} = 13\TeV$} in 2015 and 2016 at
  {CMS}'',} \textit{ Eur. Phys. J. C} \textbf{ 81} (2021) 800,
  \href{http://dx.doi.org/10.1140/epjc/s10052-021-09538-2}{\doi{10.1140/epjc/s10052-021-09538-2}},
  \href{http://www.arXiv.org/abs/2104.01927}{\texttt{arXiv:2104.01927}}.

\bibitem{CMS-PAS-LUM-17-004}
\href {https://cds.cern.ch/record/2621960}{{CMS Collaboration}, ``{CMS}
  luminosity measurements for the 2017 data taking period'',} Technical Report
  CMS-PAS-LUM-17-004, CERN, 2017.

\bibitem{CMS-PAS-LUM-18-002}
\href {https://cds.cern.ch/record/2676164}{{CMS Collaboration}, ``{CMS}
  luminosity measurements for the 2018 data taking period'',} Technical Report
  CMS-PAS-LUM-18-002, CERN, 2018.

\bibitem{1126-6708-2004-11-040}
\hrefCMSnoop {}{P.~Nason, ``{A new method for combining NLO QCD with shower
  Monte Carlo algorithms}'',} \textit{ JHEP} \textbf{ 11} (2004) 040,
  \href{http://dx.doi.org/10.1088/1126-6708/2004/11/040}{\doi{10.1088/1126-6708/2004/11/040}},
  \href{http://www.arXiv.org/abs/hep-ph/0409146}{\texttt{arXiv:hep-ph/0409146}}.

\bibitem{1126-6708-2007-11-070}
\hrefCMSnoop {}{S.~Frixione, P.~Nason, and C.~Oleari, ``{Matching NLO QCD
  computations with parton shower simulations: the POWHEG method}'',} \textit{
  JHEP} \textbf{ 11} (2007) 070,
  \href{http://dx.doi.org/10.1088/1126-6708/2007/11/070}{\doi{10.1088/1126-6708/2007/11/070}},
  \href{http://www.arXiv.org/abs/0709.2092}{\texttt{arXiv:0709.2092}}.

\bibitem{Alioli2010}
\hrefCMSnoop {}{S.~Alioli, P.~Nason, C.~Oleari, and E.~Re, ``{A general
  framework for implementing NLO calculations in shower Monte Carlo programs:
  the \POWHEG BOX}'',} \textit{ JHEP} \textbf{ 06} (2010) 043,
  \href{http://dx.doi.org/10.1007/JHEP06(2010)043}{\doi{10.1007/JHEP06(2010)043}},
  \href{http://www.arXiv.org/abs/1002.2581}{\texttt{arXiv:1002.2581}}.

\bibitem{1126-6708-2007-09-126}
\hrefCMSnoop {}{S.~Frixione, P.~Nason, and G.~Ridolfi, ``{A positive-weight
  next-to-leading-order Monte Carlo for heavy flavour hadroproduction}'',}
  \textit{ JHEP} \textbf{ 09} (2007) 126,
  \href{http://dx.doi.org/10.1088/1126-6708/2007/09/126}{\doi{10.1088/1126-6708/2007/09/126}},
  \href{http://www.arXiv.org/abs/0707.3088}{\texttt{arXiv:0707.3088}}.

\bibitem{hartanto_higgs_2015}
\hrefCMSnoop {}{H.~B. Hartanto, B.~Jager, L.~Reina, and D.~Wackeroth, ``{Higgs}
  boson production in association with top quarks in the {\POWHEG} {BOX}'',}
  \textit{ Phys. Rev. D} \textbf{ 91} (2015) 094003,
  \href{http://dx.doi.org/10.1103/PhysRevD.91.094003}{\doi{10.1103/PhysRevD.91.094003}},
  \href{http://www.arXiv.org/abs/1501.04498}{\texttt{arXiv:1501.04498}}.

\bibitem{luisoni_hwhz_2013}
\hrefCMSnoop {}{G.~Luisoni, P.~Nason, C.~Oleari, and F.~Tramontano,
  ``{$\PH\PW^{\pm}$/$\PH\PZ$} + 0 and 1 jet at {NLO} with the {\POWHEG} {BOX}
  interfaced to {GoSam} and their merging within {MiNLO}'',} \textit{ JHEP}
  \textbf{ 10} (2013) 083,
  \href{http://dx.doi.org/10.1007/JHEP10(2013)083}{\doi{10.1007/JHEP10(2013)083}},
  \href{http://www.arXiv.org/abs/1306.2542}{\texttt{arXiv:1306.2542}}.

\bibitem{kolmogorov}
\hrefCMSnoop {}{A.~N. Kolmogorov, ``Sulla determinazione empirica di una legge
  di distribuzione'',} \textit{ Giornale dell'Istituto Italiano degli Attuari}
  \textbf{ 4} (1933) 83.

\bibitem{smirnov1948}
\hrefCMSnoop {}{N.~Smirnov, ``Table for estimating the goodness of fit of
  empirical distributions'',} \textit{ Ann. Math. Statist.} \textbf{ 19} (1948)
  279,
  \href{http://dx.doi.org/10.1214/aoms/1177730256}{\doi{10.1214/aoms/1177730256}}.

\bibitem{James:1019859}
F.~E. James, ``Statistical Methods in Experimental Physics; 2nd ed.''.
\newblock World Scientific, Singapore, 2006.
\newblock \href{http://dx.doi.org/10.1142/6096}{\doi{10.1142/6096}},
  ISBN~978-981-270-527-3.

\bibitem{Butterworth:2015oua}
\hrefCMSnoop {}{J.~Butterworth { et~al.}, ``{PDF4LHC recommendations for LHC
  Run II}'',} \textit{ J. Phys. G} \textbf{ 43} (2016) 023001,
  \href{http://dx.doi.org/10.1088/0954-3899/43/2/023001}{\doi{10.1088/0954-3899/43/2/023001}},
  \href{http://www.arXiv.org/abs/1510.03865}{\texttt{arXiv:1510.03865}}.

\bibitem{ppInelastic2018}
\hrefCMSnoop {}{{CMS Collaboration}, ``{Measurement of the inelastic
  proton-proton cross section at $\sqrt{s}=13\TeV$}'',} \textit{ JHEP} \textbf{
  07} (2018) 161,
  \href{http://dx.doi.org/10.1007/JHEP07(2018)161}{\doi{10.1007/JHEP07(2018)161}},
  \href{http://www.arXiv.org/abs/1802.02613}{\texttt{arXiv:1802.02613}}.

\bibitem{pvalue}
\hrefCMSnoop {}{L.~Demortier, ``{P} values and nuisance parameters'',} in
  \textit{ Statistical issues for {LHC} physics. {Proc. Workshop on statistical
  issues for LHC physics, PHYSTAT-LHC, Geneva, Switzerland, June} 27--29,
  2007}, p.~23.
\newblock 2008.
\newblock
  \href{http://dx.doi.org/10.5170/CERN-2008-001}{\doi{10.5170/CERN-2008-001}}.

\bibitem{CLS2}
\hrefCMSnoop {}{T.~Junk, ``{Confidence level computation for combining searches
  with small statistics}'',} \textit{ Nucl. Instrum. Meth. A} \textbf{ 434}
  (1999) 435,
  \href{http://dx.doi.org/10.1016/S0168-9002(99)00498-2}{\doi{10.1016/S0168-9002(99)00498-2}},
  \href{http://www.arXiv.org/abs/hep-ex/9902006}{\texttt{arXiv:hep-ex/9902006}}.

\bibitem{CLS1}
\hrefCMSnoop {}{A.~L. Read, ``{Presentation of search results: The \CLs
  technique}'',} \textit{ J. Phys. G} \textbf{ 28} (2002) 2693,
  \href{http://dx.doi.org/10.1088/0954-3899/28/10/313}{\doi{10.1088/0954-3899/28/10/313}}.

\bibitem{LHCCLs}
\href {https://cds.cern.ch/record/1379837}{{ATLAS and CMS Collaborations, and
  LHC Higgs Combination Group}, ``Procedure for the {LHC} {Higgs} boson search
  combination in {Summer} 2011'',} Technical Report CMS-NOTE-2011-005,
  ATL-PHYS-PUB-2011-11, 2011.

\bibitem{AsympCLs}
\hrefCMSnoop {}{G.~Cowan, K.~Cranmer, E.~Gross, and O.~Vitells, ``Asymptotic
  formulae for likelihood-based tests of new physics'',} \textit{ Eur. Phys. J.
  C} \textbf{ 71} (2011) 1554,
  \href{http://dx.doi.org/10.1140/epjc/s10052-011-1554-0}{\doi{10.1140/epjc/s10052-011-1554-0}},
  \href{http://www.arXiv.org/abs/1007.1727}{\texttt{arXiv:1007.1727}}.
  [Erratum: \DOI{10.1140/epjc/s10052-013-2501-z}].

\bibitem{hepdata}
\hrefCMSnoop {}{}{HEPD}ata record for this analysis, 2024.
\newblock
  \href{http://dx.doi.org/10.17182/hepdata.144172}{\doi{10.17182/hepdata.144172}}.

\end{thebibliography}\endgroup
\cleardoublepage \appendix\section{The CMS Collaboration \label{app:collab}}\begin{sloppypar}\hyphenpenalty=5000\widowpenalty=500\clubpenalty=5000
\cmsinstitute{Yerevan Physics Institute, Yerevan, Armenia}
{\tolerance=6000
A.~Hayrapetyan, A.~Tumasyan\cmsAuthorMark{1}\cmsorcid{0009-0000-0684-6742}
\par}
\cmsinstitute{Institut f\"{u}r Hochenergiephysik, Vienna, Austria}
{\tolerance=6000
W.~Adam\cmsorcid{0000-0001-9099-4341}, J.W.~Andrejkovic, T.~Bergauer\cmsorcid{0000-0002-5786-0293}, S.~Chatterjee\cmsorcid{0000-0003-2660-0349}, K.~Damanakis\cmsorcid{0000-0001-5389-2872}, M.~Dragicevic\cmsorcid{0000-0003-1967-6783}, P.S.~Hussain\cmsorcid{0000-0002-4825-5278}, M.~Jeitler\cmsAuthorMark{2}\cmsorcid{0000-0002-5141-9560}, N.~Krammer\cmsorcid{0000-0002-0548-0985}, A.~Li\cmsorcid{0000-0002-4547-116X}, D.~Liko\cmsorcid{0000-0002-3380-473X}, I.~Mikulec\cmsorcid{0000-0003-0385-2746}, J.~Schieck\cmsAuthorMark{2}\cmsorcid{0000-0002-1058-8093}, R.~Sch\"{o}fbeck\cmsorcid{0000-0002-2332-8784}, D.~Schwarz\cmsorcid{0000-0002-3821-7331}, M.~Sonawane\cmsorcid{0000-0003-0510-7010}, S.~Templ\cmsorcid{0000-0003-3137-5692}, W.~Waltenberger\cmsorcid{0000-0002-6215-7228}, C.-E.~Wulz\cmsAuthorMark{2}\cmsorcid{0000-0001-9226-5812}
% ...
\par}
\vskip\cmsinstskip
\dag:~Deceased\\
$^{1}$Also at Yerevan State University, Yerevan, Armenia\\
$^{2}$Also at TU Wien, Vienna, Austria\\
% ...
\end{sloppypar}
\end{document}